%% file: A_main_arxiv_v3.tex
\documentclass[journal,english]{IEEEtran}
\usepackage{amsmath,amsfonts, amssymb}
\usepackage{algcompatible}
\usepackage{array}
\usepackage[caption=false,font=normalsize,labelfont=sf,textfont=sf]{subfig}
\usepackage{textcomp}
\usepackage{stfloats}
\usepackage{url}
\usepackage{verbatim}
\usepackage{xcolor}
\usepackage[normalem]{ulem}
\usepackage{graphicx}
\usepackage{cite}
\usepackage{pstricks}
\usepackage{comment}
\usepackage{booktabs} 
\usepackage{bm}   
\usepackage{amsthm}
\DeclareMathOperator{\yarg}{arg}
\DeclareMathOperator{\Tr}{Tr}

\theoremstyle{plain} 
\newtheorem{theorem}{Theorem}

\newtheorem{remark}{Remark}
\usepackage{algorithm}
\usepackage{algpseudocode}

\input{macros}

\usepackage{microtype}

\def\BibTeX{{\rm B\kern-.05em{\sc i\kern-.025em b}\kern-.08em
		T\kern-.1667em\lower.7ex\hbox{E}\kern-.125emX}}
\usepackage{hyperref}
\usepackage{balance}
\hypersetup{
	colorlinks=true,   
	linkcolor=black,   
	citecolor=black, 
	urlcolor=blue,     
	hidelinks        
}
\usepackage[acronym,nomain]{glossaries}
\makeglossaries

\newacronym{bs}{BS}{base station}
\newacronym{ris}{RIS}{reconfigurable intelligent surface}
\newacronym{ue}{UE}{user equipment}
\newacronym{eve}{Eve}{eavesdropper}

\newacronym{mimo}{MIMO}{multiple-input multiple-output}
\newacronym{miso}{MISO}{multiple-input single-output}
\newacronym{simo}{SIMO}{single-input multiple-output}
\newacronym{csi}{CSI}{channel state information}
\newacronym{sinr}{SINR}{signal-to-interference-plus-noise ratio}
\newacronym{snr}{SNR}{signal-to-noise ratio}
\newacronym{mm}{MM}{majorization-minimization}

\newacronym{bcd}{BCD}{block coordinate descent}
\newacronym{gn}{GN}{Gaussian noise}
\newacronym{ej}{EJ}{encoded jamming}
\newacronym{jdg}{JDG}{jamming discrimination gain}

\newacronym{wmmse}{WMMSE}{weighted minimum mean-square error}
\newacronym{ao}{AO}{alternating optimization}
\newacronym{kkt}{KKT}{Karush-Kuhn-Tucker}
\newacronym{lln}{LLN}{law of large numbers}
\begin{document}
	\title{Encoded Jamming Secure Communication for RIS-Assisted Systems}
	\author{Hao~Yang, Hao~Xu,~\IEEEmembership{Senior Member,~IEEE,}  Kai~Wan,~\IEEEmembership{Member,~IEEE,} Sijie~Zhao, and Robert~Caiming~Qiu,~\IEEEmembership{Fellow,~IEEE} 
		
 \thanks{
		A short version of this paper   was accepted by     the 2026 IEEE/CIC International Conference on Communications in China (ICCC). 
	}	
		\thanks{H.~Yang,  K.~Wan, S.~Zhao, and R.~C.~Qiu are with the School of Electronic Information and Communications,
			Huazhong University of Science and Technology, 430074  Wuhan, China,  (e-mail: \{hao\_yang, kai\_wan,zhaosijie,caiming\}@hust.edu.cn).}
		\thanks{H.~Xu is with the National Mobile Communications Research Laboratory, Southeast University, Nanjing 210096, China (e-mail: hao.xu@seu.edu.cn).}
	}
	\maketitle
	\begin{abstract}
		This paper investigates a cooperative jamming (CJ)-aided secure wireless communication system. Conventional CJ schemes transmit Gaussian noise (GN) to improve security, which inherently degrades the legitimate receiver's performance. While the encoded jamming (EJ) scheme mitigates this interference, its advantage over the GN scheme is highly channel-dependent. To address this limitation, we introduce a reconfigurable intelligent surface (RIS)-assisted secure design based on the EJ scheme for secrecy-rate maximization. Starting from the multiple-input multiple-output (MIMO) secrecy-rate expressions for the EJ and GN schemes, we identify the jamming-channel conditions under which the EJ scheme may be limited. For the general MIMO setup, we develop a weighted minimum mean-square error (WMMSE)-based framework for joint precoder and RIS phase optimization. We further provide a large-RIS asymptotic benchmark showing that a jamming-aligned EJ construction achieves a positive gap over a signal-aligned GN benchmark in the considered multiple-input single-output (MISO) setting. Simulation results support the analysis and show that, under the evaluated settings, the RIS-assisted EJ scheme can alleviate the identified spatial bottlenecks and improve the secrecy rate relative to the considered baselines.
	\end{abstract}
	
	
	\section{Introduction}
	
\IEEEPARstart{P}hysical layer security (PLS) has emerged as a key technology for safeguarding wireless communications, exploiting channel characteristics to achieve information-theoretic security without incurring cryptographic overhead~\cite{ref10}. However, the achievable secrecy rate, defined as the difference between the mutual information of the \gls{bs}--\gls{ue} link and that of the \gls{bs}--\gls{eve} link, is constrained by the relative quality of these links~\cite{chuRobustBeamformingTechniques2015}. To address this, artificial noise (AN) injection~\cite{wuSecureMassiveMIMO2016},~\cite{7079465} and cooperative jamming (CJ) strategies~\cite{tekin2008general,li2014secrecy} have been investigated.
	
	CJ strategies can degrade Eve's reception while maintaining legitimate communication quality. Traditional CJ schemes utilize \gls{gn} transmission from cooperative jammers~\cite{chu2014secrecy,hu2018cooperative}, but such approaches introduce unintended interference to the \gls{ue}. Various \gls{ej} schemes~\cite{xu16,xu17,xu18} employ structured interference via algebraically coded signals, offering security benefits over methods based on \gls{gn}. Specifically, \cite{xu16,xu17} analyzed the achievable secrecy rate of a discrete memoryless wiretap channel with an encoded jammer and verified the secrecy performance in a single-antenna Gaussian wiretap channel. A subsequent study~\cite{xu18} investigated a similar scalar Gaussian wiretap channel, demonstrating that lattice-structured codes prevent the achievable secrecy rate from saturating at high signal-to-noise ratios.
For the Gaussian \gls{mimo} wiretap channel, \cite{xu2024} proposed an \gls{ej} scheme allowing the jammer to switch between the \gls{gn} and \gls{ej} schemes. This strategy formulates the problem as a special case of the two-user wiretap channel, where both users transmit secret messages using Gaussian random coding and lattice-based codes. Comparisons in~\cite{xu2024} revealed that the \gls{ej} scheme does not consistently achieve higher secrecy rates than the \gls{gn} scheme. Specifically, the performance gain of the \gls{ej} scheme diminishes when the jammer-to-Eve channel is much stronger than the jammer-to-UE channel, or when the jammer's transmit power is not dominant.
	
	As a key technology for 6G, \gls{ris} offers programmable control over electromagnetic wave propagation for dynamic signal redirection and precise beamforming~\cite{wuIntelligentReflectingSurface2018,direnzoSmartRadioEnvironments2020}. An RIS consists of a large array of low-cost reflecting elements, each capable of imposing an adjustable phase shift. Consequently, an RIS can mitigate growing security threats from eavesdroppers by intelligently manipulating wireless channel conditions~\cite{kaurSurveyReconfigurableIntelligent2024}.
	A substantial body of research has explored RIS-assisted secure wireless communications~\cite{cuiSecureWirelessCommunication2019,illiEnhancingPhysicalLayer2024,yuEnablingSecureWireless2019,guanIntelligentReflectingSurface2020a,arzykulovArtificialNoiseRISaided2023,zhangPhysicalLayerSecurity2021,RISMIMO1,wuMIMOSecureCommunication2022,wangEnergyEfficientRobust2021,chenPhysicalLayerSecurity2024}. In particular, \cite{cuiSecureWirelessCommunication2019,yuEnablingSecureWireless2019} considered systems with multi-antenna transmitters and single-antenna users and eavesdroppers. Through the joint optimization of transmit beamforming and the RIS phase shift configuration, these works demonstrated that an RIS can simultaneously enhance legitimate links and suppress eavesdropping. \cite{arzykulovArtificialNoiseRISaided2023} proposed an RIS element partitioning strategy, dividing the surface units into signal-enhancement and AN-enhancement groups; by jointly optimizing the partition ratio and power allocation, this approach significantly improves the secrecy capacity. Furthermore, Guan \emph{et al.}~\cite{guanIntelligentReflectingSurface2020a} analyzed RIS-assisted \gls{miso} networks and demonstrated that the joint optimization of transmit beamforming and the RIS phase shift matrix improves energy efficiency under perfect \gls{csi}.
For MIMO scenarios, \cite{zhangPhysicalLayerSecurity2021} employed stochastic geometry to analyze security performance with randomly distributed users, showing that more RIS elements substantially reduce the secrecy outage probability. \cite{RISMIMO1} proposed a secure MIMO system assisted by an RIS and enhanced with AN. Employing \gls{bcd} and \gls{mm} algorithms, the authors jointly optimized the transmit precoder and RIS phase shift configuration to maximize the secrecy rate. In~\cite{wuMIMOSecureCommunication2022}, the authors demonstrated that jointly optimizing the precoder and RIS phase shifts provides security improvements even under finite phase-shift resolution constraints.
	
	RIS-assisted CJ research mainly focuses on MISO networks. Here, the \gls{gn} scheme is the cooperative-jammer counterpart of transmitter-side AN: both use non-decodable Gaussian interference, but the former has separate power and channel constraints. Wang \emph{et al.}~\cite{wangEnergyEfficientRobust2021} investigated robust joint beamforming and jamming under imperfect \gls{csi}. Deep reinforcement learning was adopted to optimize the RIS phase shift configurations and CJ signals in~\cite{zhang2023}. Additionally, \cite{Wen2025} examined RIS-assisted CJ in symbiotic radio scenarios, enhancing both secrecy and spectral efficiency. While \gls{ris} has been extensively integrated with conventional \gls{gn} or AN schemes to enhance physical layer security~\cite{wangEnergyEfficientRobust2021,illiEnhancingPhysicalLayer2024}, its combination with the \gls{ej} scheme has received little attention. This gap is particularly relevant because the relative performance of the \gls{ej} and \gls{gn} schemes is strongly affected by the effective information and jamming channels. Unlike transmit precoding alone, an \gls{ris} provides an additional spatial degree of freedom for jointly reshaping the BS--UE, BS--Eve, jammer--UE, and jammer--Eve links. Motivated by this, we investigate whether RIS-assisted channel reconfiguration can alleviate unfavorable jamming-channel conditions and improve the secrecy performance of both schemes.

	Although the \gls{ej} scheme enhances physical layer security by mitigating the interference caused by the conventional \gls{gn} scheme, it does not consistently achieve higher secrecy rates than the \gls{gn} scheme across all channel conditions (see Theorem~\ref{thm:mi_bounds}). Furthermore, applying the \gls{ej} scheme to \gls{ris}-assisted MIMO systems introduces optimization challenges. Specifically, the \gls{ue} must decode and cancel the jamming signal, whereas the eavesdropper must be unable to decode it despite having full knowledge of the codebooks. This dual requirement translates the secrecy rate maximization into a non-convex max-min optimization problem.
	To the best of our knowledge, the only existing algorithm for the \gls{ej} scheme in MIMO systems was recently proposed in~\cite{xu2024}, relying on the simultaneous diagonalization (SD) of Hermitian matrices. However, the SD method in~\cite{xu2024} is developed for fixed MIMO channel matrices and therefore does not directly address the joint optimization of the transmit precoders and RIS phase shifts in the RIS-assisted setting, where the same RIS phase shifts simultaneously affect the BS--UE, BS--Eve, jammer--UE, and jammer--Eve links. To address these limitations and the channel-dependent performance bottleneck of the \gls{ej} scheme, we propose a joint transmit precoding and \gls{ris} phase shift optimization framework. The main objective of this paper is to clarify the channel-dependent performance relationship between the \gls{ej} and \gls{gn} schemes, determine the extent to which introducing an \gls{ris} can improve the secrecy performance of both schemes and change their relative performance, and establish a theoretical characterization of that relationship in the large-\gls{ris} regime. To achieve this objective, the main contributions of this paper are summarized as follows:

	$\bullet$ \textit{Jamming-channel bottleneck analysis:}
		Xu \emph{et al.}~\cite{xu2024} observed that the ordering of the \gls{ej} and \gls{gn} schemes depends on channel conditions. However, a general MIMO sufficient condition identifying the channel structure responsible for an \gls{ej} loss relative to \gls{gn} was not established. We identify the mechanism by separately analyzing the two rate constraints in the \gls{ej} achievable-rate expression and combining mutual-information identities with a matrix partial-order analysis. This gives a spatial degradedness condition under which the sum-rate constraint limits the \gls{ej} rate, explaining why the \gls{ej} scheme can be weaker than the \gls{gn} scheme.
	
	$\bullet$ \textit{RIS-assisted MIMO optimization and EJ-WMMSE:}
		The lower-bound construction used to handle the non-smooth EJ objective follows~\cite{xu2024}. Adding the \gls{ris} makes each subproblem more involved because the same RIS phase shifts simultaneously affect the BS--UE, BS--Eve, jammer--UE, and jammer--Eve links, which requires joint optimization of the transmit precoders and RIS phases. We derive WMMSE-based BCD updates for the transmit precoders and MM updates for the \gls{ris} phases, and evaluate the resulting candidates using the original EJ secrecy-rate expression. The proposed algorithm also achieves higher secrecy rates than the SD algorithm in the evaluated RIS-free MIMO settings.
	
	$\bullet$ \textit{Large-\gls{ris} asymptotic characterization:}
		For the MISO strong-jamming benchmark, we construct signal-aligned and jamming-aligned phase configurations and use the strong law of large numbers, the central limit theorem, and the continuous mapping theorem to distinguish the quadratic coherent power gain of the aligned link from the linear non-coherent scaling of the remaining links. These scaling laws yield explicit transmit-power and jamming-ratio thresholds under which the jamming-aligned \gls{ej} construction achieves a positive asymptotic ergodic gap over the signal-aligned \gls{gn} benchmark. For the SIMO case, the same order separation establishes an \gls{ris} phase configuration that reverses the jamming-channel dominance with probability approaching one.

\textit{Paper organization:} Section~\ref{sec:mi_relationship} reviews the secrecy-rate expressions for the \gls{ej} and \gls{gn} schemes and identifies the jamming-channel bottleneck. Section~\ref{smpf} incorporates these expressions into an RIS-assisted MIMO channel model and formulates the secrecy-rate maximization problems. Section~\ref{sec:algorithm} develops the EJ-WMMSE algorithm. Section~\ref{sec:asymptotic} presents the large-RIS asymptotic benchmark analysis. Section~\ref{simu} presents simulation results, and Section~\ref{conclusion} concludes the paper.
	
\textit{Notation}: $\mathbb{C}$ denotes the complex field. Boldface lower- and upper-case letters denote vectors and matrices, respectively. $\mathbf{I}_N$ is the $N \times N$ identity matrix, and $\mathbf{0}$ denotes an all-zero vector or matrix. $[\mathbf{A}]_{i,j}$ and $[\mathbf{a}]_n$ denote the $(i,j)$-th entry of $\mathbf{A}$ and the $n$-th entry of $\mathbf{a}$, respectively. Superscripts $(\cdot)^{\mathrm{T}}$, $(\cdot)^*$, and $(\cdot)^{\mathrm{H}}$ denote transpose, conjugate, and conjugate transpose, respectively. $\mathrm{Tr}(\cdot)$ and $|\cdot|$ denote matrix trace and determinant, respectively. $\mathbb{E}[\cdot]$ denotes statistical expectation. $\mathrm{diag}(\mathbf{x})$ denotes a diagonal matrix with $\mathbf{x}$ on its main diagonal. $\mathbf{A} \succ 0$ ($\mathbf{A} \succeq 0$) indicates that $\mathbf{A}$ is positive definite (positive semidefinite). $\|\cdot\|$ denotes the Euclidean norm of a vector or the Frobenius norm of a matrix. Moreover, $[\cdot]^+ \triangleq \max(\cdot,0)$, $\odot$ denotes the Hadamard product, and $\log(\cdot)$ denotes the natural logarithm.\footnote{Natural-log rate expressions used in the optimization derivation are converted to bits/s/Hz by the factor $1/\ln 2$ when reporting numerical results, whereas $\log_2(\cdot)$ is written explicitly in the asymptotic analysis.} $\mathcal{O}(\cdot)$ denotes deterministic Big-$\mathcal{O}$ notation, i.e., $f(M)=\mathcal{O}(g(M))$ implies $\limsup_{M\to\infty}|f(M)|/g(M)<\infty$. $\mathcal{O}_p(\cdot)$ denotes asymptotic order in probability. Specifically, $X_M=\mathcal{O}_p(a_M)$ as $M\to\infty$ means that, for any $\epsilon>0$, there exist a finite constant $C>0$ and a positive integer $M_0$ such that $\Pr[|X_M/a_M|>C]<\epsilon$ for all $M>M_0$.

		\section{CJ-Aided MIMO Wiretap Model and Jamming-Channel Bottleneck}
		\label{sec:mi_relationship}
		
		Before introducing the RIS-assisted channel structure, we first recall the achievable secrecy-rate expressions for the \gls{ej} and \gls{gn} schemes in a CJ-aided MIMO wiretap channel~\cite{xu2024}. Building on these expressions, we then derive structural relationships between the two schemes and identify the jamming-channel bottleneck that motivates the RIS-assisted design.

		\subsection{CJ-Aided MIMO Model and Achievable Secrecy Rates}
		The transmitted complex baseband signals are modeled as $\mathbf{x}_1 = \mathbf{F}_1 \mathbf{s}_1$ and $\mathbf{x}_2 = \mathbf{F}_2 \mathbf{s}_2$,
		where $\mathbf{s}_1 \sim \mathcal{CN}(\mathbf{0}, \mathbf{I}_{N_{\mathrm{b}} })$ and $\mathbf{s}_2 \sim \mathcal{CN}(\mathbf{0}, \mathbf{I}_{N_{\mathrm{c}}})$ denote the legitimate information and jamming signal vectors, respectively. The corresponding precoding matrices are $\mathbf{F}_1 \in \mathbb{C}^{N_{\mathrm{b}} \times N_{\mathrm{b}} }$ and $\mathbf{F}_2 \in \mathbb{C}^{N_{\mathrm{c}} \times N_{\mathrm{c}}}$, with covariance matrices $\mathbf{Q}_1=\mathbf{F}_1\mathbf{F}_1^{\mathrm{H}}$ and $\mathbf{Q}_2=\mathbf{F}_2\mathbf{F}_2^{\mathrm{H}}$. The received signals at the UE and Eve are
		\begin{align}
			\mathbf{y}_{\mathrm{u}} &= \mathbf{H}_1\mathbf{F}_1\mathbf{s}_1
			+ \mathbf{H}_2\mathbf{F}_2\mathbf{s}_2
			+ \mathbf{n}_{\mathrm{u}},\\
			\mathbf{y}_{\mathrm{e}} &= \mathbf{G}_1\mathbf{F}_1\mathbf{s}_1
			+ \mathbf{G}_2\mathbf{F}_2\mathbf{s}_2
			+ \mathbf{n}_{\mathrm{e}},
		\end{align}
where \(\mathbf{H}_1 \in \mathbb{C}^{N_{\mathrm{u}}\times N_{\mathrm{b}}}\) and \(\mathbf{H}_2 \in \mathbb{C}^{N_{\mathrm{u}}\times N_{\mathrm{c}}}\) denote the BS--UE and jammer--UE channels, respectively, and \(\mathbf{G}_1 \in \mathbb{C}^{N_{\mathrm{e}}\times N_{\mathrm{b}}}\) and \(\mathbf{G}_2 \in \mathbb{C}^{N_{\mathrm{e}}\times N_{\mathrm{c}}}\) the corresponding channels to Eve. \(N_{\mathrm{b}}\), \(N_{\mathrm{c}}\), \(N_{\mathrm{u}}\), and \(N_{\mathrm{e}}\) are the numbers of antennas at the BS, jammer, UE, and Eve, respectively. The noise vectors satisfy \(\mathbf{n}_{\mathrm{u}} \sim \mathcal{CN}(\mathbf{0}, \mathbf{I}_{N_{\mathrm{u}}})\) and \(\mathbf{n}_{\mathrm{e}} \sim \mathcal{CN}(\mathbf{0}, \mathbf{I}_{N_{\mathrm{e}}})\).
		
		For given \(\mathbf{Q}_1\) and \(\mathbf{Q}_2\), if the jammer uses the \gls{ej} scheme, an achievable secrecy rate under the strong secrecy metric satisfies
		\(R \le R_{\mathrm{EJ}}=\max\{\min\{\hat{R},\tilde{R}\},\bar{R}\}\)~\cite{xu2024},
		where
		\begin{equation} \label{SR_hat_tilde_bar_MIMO}
			\begin{aligned}
				\hat{R} =& \biggl[ \log\left|\mathbf{H}_1 \mathbf{Q}_1 \mathbf{H}_1^{\mathrm{H}} + \mathbf{I}_{N_{\mathrm{u}}}\right|\\
				- &\log\left|\mathbf{G}_1 \mathbf{Q}_1 \mathbf{G}_1^{\mathrm{H}} \left(\mathbf{G}_2 \mathbf{Q}_2 \mathbf{G}_2^{\mathrm{H}} + \mathbf{I}_{N_{\mathrm{e}}}\right)^{-1} + \mathbf{I}_{N_{\mathrm{e}}}\right|
				\biggr]^+,\\
				\tilde{R} =& \biggl[ \log\left|\mathbf{H}_1 \mathbf{Q}_1 \mathbf{H}_1^{\mathrm{H}} + \mathbf{H}_2 \mathbf{Q}_2 \mathbf{H}_2^{\mathrm{H}} + \mathbf{I}_{N_{\mathrm{u}}}\right| \\
				-&\log\left|\mathbf{G}_1 \mathbf{Q}_1 \mathbf{G}_1^{\mathrm{H}} + \mathbf{G}_2 \mathbf{Q}_2 \mathbf{G}_2^{\mathrm{H}} + \mathbf{I}_{N_{\mathrm{e}}}\right|
				\biggr]^+,\\
				\bar{R} = &\bigl[ \log\left|\mathbf{H}_1 \mathbf{Q}_1 \mathbf{H}_1^{\mathrm{H}} + \mathbf{I}_{N_{\mathrm{u}}}\right|
				- \log\left|\mathbf{G}_1 \mathbf{Q}_1 \mathbf{G}_1^{\mathrm{H}} + \mathbf{I}_{N_{\mathrm{e}}}\right|
				\bigr]^+.
			\end{aligned}
		\end{equation}
In the encoded-jamming framework, $\min\{\hat{R}, \tilde{R}\}$ corresponds to the secure coding scheme for the discrete memoryless multiple-access wiretap channel developed in~\cite{yassaeeMultipleAccessWiretap2010, xu2023achievable}. $\bar{R}$ represents the trivial case where the jammer transmits zero power. For the GN scheme, the achievable secrecy rate under the same CJ-aided MIMO formulation is given by
		\begin{equation}
			\begin{aligned}
				R_{\mathrm{GN}} &= \biggl[ \log \left|\mathbf{H}_1 \mathbf{Q}_1 \mathbf{H}_1^{\mathrm{H}} \left(\mathbf{H}_2 \mathbf{Q}_2 \mathbf{H}_2^{\mathrm{H}} + \mathbf{I}_{N_{\mathrm{u}}}\right)^{-1} + \mathbf{I}_{N_{\mathrm{u}}} \right|\\
				&-\log\left|\mathbf{G}_1 \mathbf{Q}_1 \mathbf{G}_1^{\mathrm{H}} \left(\mathbf{G}_2 \mathbf{Q}_2 \mathbf{G}_2^{\mathrm{H}} + \mathbf{I}_{N_{\mathrm{e}}}\right)^{-1} + \mathbf{I}_{N_{\mathrm{e}}}\right|
				\biggr]^+.
				\label{eq:GN_rate}
			\end{aligned}
		\end{equation}	
\subsection{Jamming-Channel Bottleneck}	
		Based on the definitions above, we analyze the performance boundaries of the \gls{ej} scheme and demonstrate the limitations of relying solely on transmit precoding.  
		With a slight abuse of notation, \(\hat{R}\), \(\tilde{R}\), and \(R_{\mathrm{GN}}\) in this subsection denote the corresponding untruncated expressions obtained by removing \([\cdot]^+\) from \eqref{SR_hat_tilde_bar_MIMO} and \eqref{eq:GN_rate}. 

		\begin{theorem}\label{thm:mi_bounds}
			For any channel realization and transmit covariance matrices \(\mathbf{Q}_1, \mathbf{Q}_2 \succeq 0\), the following hold:
		
				1) Relation between $\hat{R}$ and $R_{\mathrm{GN}}$: $\hat{R} \ge R_{\mathrm{GN}}$.
				
				2) Relation between $\tilde{R}$ and $R_{\mathrm{GN}}$: Their difference is determined by the \gls{jdg} $\Delta_{\mathrm{jam}}$, yielding $\tilde{R} - R_{\mathrm{GN}} \triangleq \Delta_{\mathrm{jam}}$, where $\Delta_{\mathrm{jam}} = \log\big|\mathbf{I}_{N_{\mathrm{c}}}+\mathbf{F}_2^{\mathrm{H}}\mathbf{H}_2^{\mathrm{H}}\mathbf{H}_2\mathbf{F}_2\big| - \log\big|\mathbf{I}_{N_{\mathrm{c}}}+\mathbf{F}_2^{\mathrm{H}}\mathbf{G}_2^{\mathrm{H}}\mathbf{G}_2\mathbf{F}_2\big|$.
				
				3) Spatial degradedness bottleneck: If Eve's jamming channel spatially dominates the UE's jamming channel such that
				\(
				\mathbf{G}_2^{\mathrm{H}}\mathbf{G}_2 \succeq \mathbf{H}_2^{\mathrm{H}}\mathbf{H}_2,
				\)
				then \(\hat{R} \ge \tilde{R}\) for any feasible precoding matrix. {Consequently, $\tilde{R}$ determines $\min\{\hat{R},\tilde{R}\}$, resulting in a non-positive \gls{jdg} \((\Delta_{\mathrm{jam}}\le 0)\) and}
				\(
				\min\{\hat{R},\tilde{R}\}\le R_{\mathrm{GN}}.
				\)
			
		\end{theorem}

		\begin{proof}[Proof sketch]
			Applying the chain rule of mutual information yields $\hat{R} - R_{\mathrm{GN}} = I(\mathbf{x}_1;\mathbf{y}_{\mathrm{u}}\mid \mathbf{x}_2)-I(\mathbf{x}_1;\mathbf{y}_{\mathrm{u}})$. Since $\mathbf{x}_1$ and $\mathbf{x}_2$ are independent, $I(\mathbf{x}_1;\mathbf{y}_{\mathrm{u}}\mid \mathbf{x}_2) \ge I(\mathbf{x}_1;\mathbf{y}_{\mathrm{u}})$, proving Property 1. Similarly, $\tilde{R} - R_{\mathrm{GN}} = I(\mathbf{x}_2;\mathbf{y}_{\mathrm{u}}\mid \mathbf{x}_1)-I(\mathbf{x}_2;\mathbf{y}_{\mathrm{e}}\mid \mathbf{x}_1)$. Writing the latter difference in Gaussian log-determinant form yields $\Delta_{\mathrm{jam}}$ via the Sylvester determinant identity, proving Property 2.
			For Property 3, the difference expands as $\hat{R} - \tilde{R} = \log\big|\mathbf{I}_{N_{\mathrm{c}}} + \mathbf{F}_2^{\mathrm{H}}\mathbf{G}_2^{\mathrm{H}}\mathbf{G}_2\mathbf{F}_2\big| - \log\big|\mathbf{I}_{N_{\mathrm{c}}} + \mathbf{F}_2^{\mathrm{H}}\mathbf{H}_2^{\mathrm{H}} (\mathbf{I}_{N_{\mathrm{u}}} + \mathbf{H}_1\mathbf{Q}_1\mathbf{H}_1^{\mathrm{H}})^{-1} \mathbf{H}_2\mathbf{F}_2\big|$. Since $(\mathbf{I}_{N_{\mathrm{u}}} + \mathbf{H}_1\mathbf{Q}_1\mathbf{H}_1^{\mathrm{H}})^{-1} \preceq \mathbf{I}_{N_{\mathrm{u}}}$, the condition $\mathbf{G}_2^{\mathrm{H}}\mathbf{G}_2 \succeq \mathbf{H}_2^{\mathrm{H}}\mathbf{H}_2$ implies $\mathbf{G}_2^{\mathrm{H}}\mathbf{G}_2 \succeq \mathbf{H}_2^{\mathrm{H}} (\mathbf{I}_{N_{\mathrm{u}}} + \mathbf{H}_1\mathbf{Q}_1\mathbf{H}_1^{\mathrm{H}})^{-1} \mathbf{H}_2$, which forces $\hat{R} \ge \tilde{R}$. Concurrently, this condition enforces $\Delta_{\mathrm{jam}} \le 0$, yielding $\tilde{R} \le R_{\mathrm{GN}}$. Thus, $\min\{\hat{R}, \tilde{R}\} = \tilde{R} \le R_{\mathrm{GN}}$. The detailed proof is provided in Appendix~\ref{app:mi_bounds}.
		\end{proof}
\begin{cor}[Pointwise sufficient condition for EJ dominance]\label{cor:dominance}
	For fixed channel matrices and a fixed feasible precoding strategy, if the \gls{jdg} satisfies \(\Delta_{\mathrm{jam}}\ge 0\), then the \gls{ej} rate under the same configuration satisfies \(R_{\mathrm{EJ}}\ge R_{\mathrm{GN}}\).
\end{cor}
		This condition is pointwise. It only compares the two schemes under the same fixed channel realization and the same feasible precoding strategy. It does not imply an ordering between the optimized values of the two schemes, since their respective optimization problems may select different transmit precoders. In the RIS-assisted setting introduced later, they may also select different RIS phase shifts.
		Theorem~\ref{thm:mi_bounds} highlights a property governed by the minimum-decision rule: in disadvantaged scenarios, since $\hat{R} \ge \tilde{R}$ holds, the overall secrecy rate is bounded by $\tilde{R}$. Under such severe channel degradedness ($\Delta_{\mathrm{jam}} < 0$), increasing the jamming power $P_2$ exacerbates the penalty on $\tilde{R}$, widening the performance gap in favor of the \gls{gn} scheme.
		
\begin{remark}[Spatial bottleneck in SIMO and MISO systems]\label{rem:beamforming_limitation}
			While Property 3 of Theorem~\ref{thm:mi_bounds} establishes the precoding bottleneck for general \gls{mimo} systems, it is instructive to examine how the spatial degradedness condition ($\mathbf{G}_2^{\mathrm{H}}\mathbf{G}_2 \succeq \mathbf{H}_2^{\mathrm{H}}\mathbf{H}_2$) manifests in specific antenna configurations.
		
				$\bullet$ SIMO scenario: The jammer is equipped with a single antenna, reducing $\mathbf{Q}_2$ to a scalar transmit power $P_2$, and the channels reduce to column vectors $\mathbf{h}_2$ and $\mathbf{g}_2$. The matrix partial order degenerates to a scalar gain comparison $\|\mathbf{g}_2\|^2 \ge \|\mathbf{h}_2\|^2$. Under this condition, $\Delta_{\mathrm{jam}} \le 0$ holds, and power scaling alone cannot remove this limitation for the \gls{ej} scheme.
				
				$\bullet$ \gls{miso} scenario: The channel matrices reduce to the row vectors $\mathbf{h}_2^{\mathrm{H}}$ and $\mathbf{g}_2^{\mathrm{H}}$, while the precoding matrix $\mathbf{F}_2$ reduces to a vector $\mathbf{f}_2$. For these rank-one channels, the spatial dominance condition $\mathbf{g}_2\mathbf{g}_2^{\mathrm{H}} \succeq \mathbf{h}_2\mathbf{h}_2^{\mathrm{H}}$ can hold only when the two channel vectors are collinear and Eve's channel gain is no smaller than that of the UE. Even when the channels are non-collinear, designing $\mathbf{f}_2$ such that $\Delta_{\mathrm{jam}}>0$ remains subject to two limitations. First, a spatial correlation bottleneck arises when the legitimate and eavesdropping channels are highly correlated and Eve has a substantially stronger channel gain, i.e., $\|\mathbf{g}_2\|\gg\|\mathbf{h}_2\|$. In this case, any precoding vector that delivers sufficient power to the UE for \gls{ej} decoding may induce disproportionately stronger interference at Eve, thereby tightening the joint-decoding constraint. Second, although the jammer may steer $\mathbf{f}_2$ into the null space of $\mathbf{g}_2^{\mathrm{H}}$ to ensure $\Delta_{\mathrm{jam}}>0$, doing so eliminates the jamming power received by Eve and thus undermines the primary purpose of jamming, namely, reducing the information leakage rate at Eve.
		
			Consequently, relying solely on transmit design creates a bottleneck for the \gls{ej} scheme under unfavorable channel disparities, motivating the deployment of an \gls{ris} to manipulate the channel topology.
		\end{remark}

\section{RIS-Assisted System Model and Problem Formulation}
	\label{smpf}
After introducing the generic CJ-aided MIMO wiretap formulation, we incorporate the RIS-assisted channel structure into the system model. The RIS phase shifts modify the BS--UE, jammer--UE, BS--Eve, and jammer--Eve links, and therefore affect the secrecy-rate objectives defined in Section~\ref{sec:mi_relationship}.

\begin{figure} 
		\centering
		\includegraphics[width=0.85\linewidth]{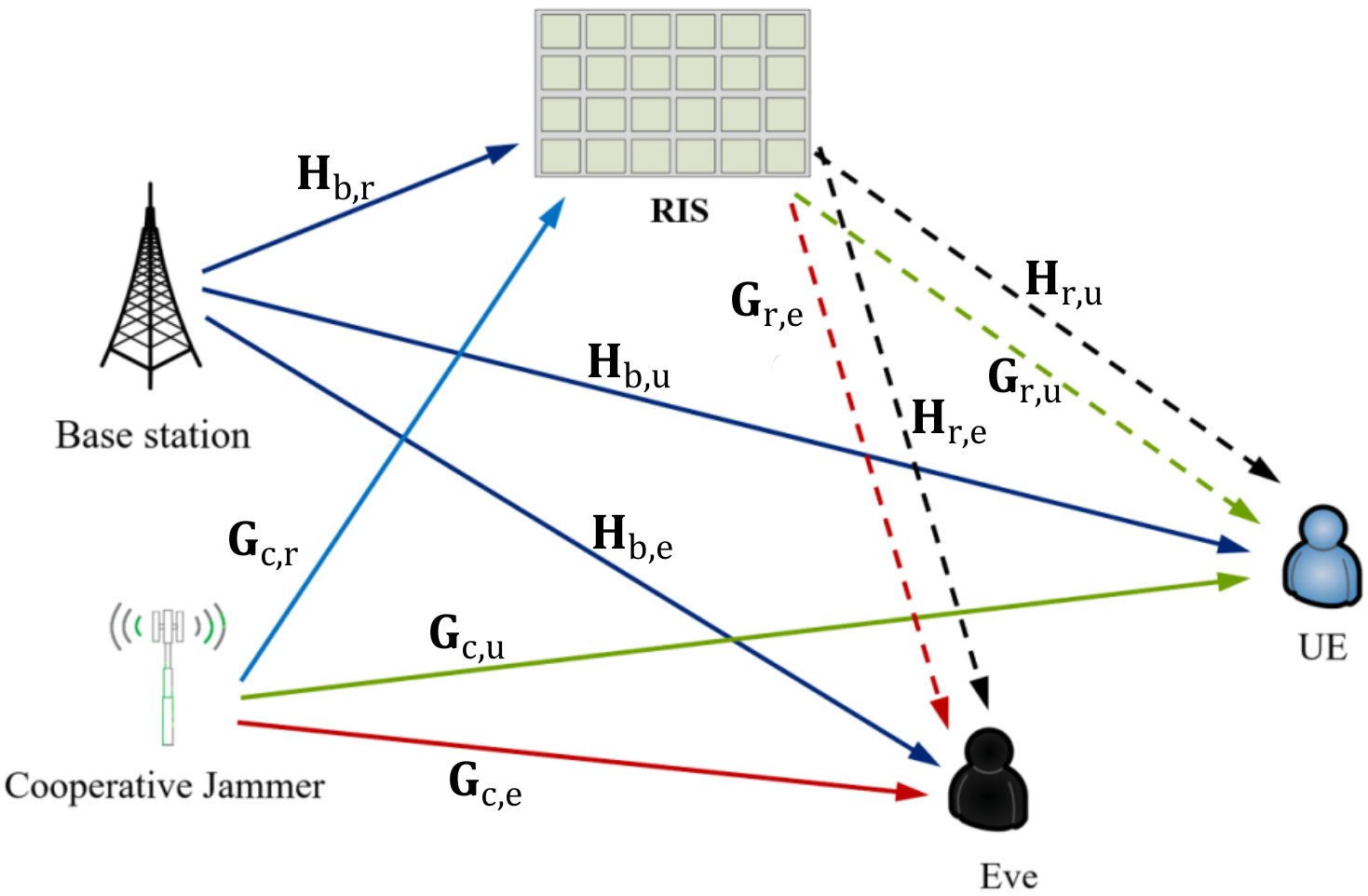}
		\caption{An RIS-assisted wireless network with a cooperative jammer.}
		\label{fig1}
	\end{figure}

	To secure communications against eavesdropping, we consider a \gls{mimo} wiretap system assisted by an \gls{ris} with $M$ reflecting elements. As shown in Fig.~\ref{fig1}, the channel matrices are denoted as follows: \(\mathbf{H}_{\mathrm{b}, \mathrm{r}} \in \mathbb{C}^{M \times N_{\mathrm{b}}}\), \(\mathbf{H}_{\mathrm{b},\mathrm{u}} \in \mathbb{C}^{N_{\mathrm{u}} \times N_{\mathrm{b}}}\), and \(\mathbf{H}_{\mathrm{b},\mathrm{e}} \in \mathbb{C}^{N_{\mathrm{e}} \times N_{\mathrm{b}}}\) represent the channels from the BS to the RIS, \gls{ue}, and Eve, respectively. Similarly, \(\mathbf{G}_{\mathrm{c},\mathrm{r}} \in \mathbb{C}^{M \times N_{\mathrm{c}}}\), \(\mathbf{G}_{\mathrm{c},\mathrm{u}} \in \mathbb{C}^{N_{\mathrm{u}} \times N_{\mathrm{c}}}\), and \(\mathbf{G}_{\mathrm{c},\mathrm{e}} \in \mathbb{C}^{N_{\mathrm{e}} \times N_{\mathrm{c}}}\) represent the channels from the jammer to the RIS, \gls{ue}, and Eve. Additionally, \(\mathbf{H}_{\mathrm{r},\mathrm{u}} \in \mathbb{C}^{N_{\mathrm{u}} \times M}\) and \(\mathbf{H}_{\mathrm{r},\mathrm{e}} \in \mathbb{C}^{N_{\mathrm{e}} \times M}\) denote the RIS-to-UE and RIS-to-Eve channels in the BS-reflected links, while \(\mathbf{G}_{\mathrm{r},\mathrm{u}} \in \mathbb{C}^{N_{\mathrm{u}} \times M}\) and \(\mathbf{G}_{\mathrm{r},\mathrm{e}} \in \mathbb{C}^{N_{\mathrm{e}} \times M}\) denote the RIS-to-UE and RIS-to-Eve channels in the jammer-reflected links. The \gls{ris} phase shift matrix is given by \(\bm{\Phi}=\mathrm{diag}\left(\bm{\phi}\right)\), where \(\bm{\phi} \triangleq [\phi_1, \dots, \phi_M]^{\mathrm{T}}\) and \(|\phi_m|=1,\,m=1,\ldots,M\). The system reduces to the SIMO case when $N_{\mathrm{b}}=N_{\mathrm{c}}=1$, and to the MISO case when $N_{\mathrm{u}}=N_{\mathrm{e}}=1$.

	The transmitted signal definitions and achievable secrecy-rate expressions follow Section~\ref{sec:mi_relationship}; this section specifies the RIS-assisted channel matrices and the resulting optimization variables.
	Under the RIS-assisted channel structure, the four link matrices in Section~\ref{sec:mi_relationship} are given by
	\begin{align}
		\mathbf{H}_1 &= \mathbf{H}_{\mathrm{b},\mathrm{u}} + \mathbf{H}_{\mathrm{r},\mathrm{u}} \bm{\Phi} \mathbf{H}_{\mathrm{b}, \mathrm{r}}, \quad
		\mathbf{H}_2 = \mathbf{G}_{\mathrm{c},\mathrm{u}}  + \mathbf{G}_{\mathrm{r},\mathrm{u}} \bm{\Phi} \mathbf{G}_{\mathrm{c},\mathrm{r}}, \nonumber   \\
		\mathbf{G}_1 &= \mathbf{H}_{\mathrm{b},\mathrm{e}} + \mathbf{H}_{\mathrm{r},\mathrm{e}} \bm{\Phi} \mathbf{H}_{\mathrm{b}, \mathrm{r}}, \quad \;
		\mathbf{G}_2 = \mathbf{G}_{\mathrm{c},\mathrm{e}}  + \mathbf{G}_{\mathrm{r},\mathrm{e}} \bm{\Phi} \mathbf{G}_{\mathrm{c},\mathrm{r}}. \nonumber
	\end{align}
	Using the RIS-assisted channel matrices and the secrecy-rate expressions in Section~\ref{sec:mi_relationship}, the secrecy-rate maximization problems for the \gls{ej} and \gls{gn} schemes are formulated as follows:
	\begin{subequations}\label{problemEJ}
		\begin{align}
			R_{\mathrm{EJ}}^{\star}= \mathop {\max }\limits_{ \Fm_{1}, \Fm_{2}, \bm{\Phi} } \quad & R_{\mathrm{EJ}}\\
			\mathrm{s.t.} \quad\; & ~ {\mathrm{Tr}}(\Fm_{1} \Fm_{1}^{\mathrm{H}}) \leq P_1,  \label{p1max}\\
			\quad\; & ~ {\mathrm{Tr}}(\Fm_{2} \Fm_{2}^{\mathrm{H}}) \leq P_2, ~\ \label{p2max}\\
			& |\phi_m| = 1,\ m = 1, \ldots ,M, \label{risris}
		\end{align}
	\end{subequations} 
	and
	\begin{align}\label{problemGN}
		R_{\mathrm{GN}}^{\star}= \mathop {\max }\limits_{ \Fm_{1}, \Fm_{2}, \bm{\Phi} } \quad & R_{\mathrm{GN}} \\
		\mathrm{s.t.} \quad\; & ~ \eqref{p1max},\eqref{p2max},\eqref{risris}. \nonumber
	\end{align}
	\section{EJ-WMMSE Algorithm for MIMO Systems}
	\label{sec:algorithm}
	
	The optimization problems in \eqref{problemEJ} and \eqref{problemGN} are non-convex due to the coupled transmit precoders, the unit-modulus RIS constraints, and the non-smooth max-min structure of the \gls{ej} scheme. This section develops a WMMSE-based algorithm for the RIS-assisted MIMO design with instantaneous CSI. The proposed algorithm is intended as a practical numerical optimizer and does not provide a closed-form optimizer for the original non-convex problem. For the \gls{ej} scheme, problem \eqref{problemEJ} is a max-min optimization problem, which is generally intractable to solve directly. As in~\cite{xu2024}, an achievable lower bound is obtained from three candidate points generated by solving the following subproblems:
	\begin{align}
		\max_{\mathbf{F}_1, \mathbf{F}_2, \bm{\Phi}} \quad & \tilde{R} \qquad \mathrm{s.t.} \quad \eqref{p1max}, \eqref{p2max}, \eqref{risris}, \label{problemrtitle} \\
		\max_{\mathbf{F}_1, \mathbf{F}_2, \bm{\Phi}} \quad & \hat{R} \qquad \mathrm{s.t.} \quad \eqref{p1max}, \eqref{p2max}, \eqref{risris}, \label{problemrhat} \\
		\max_{\mathbf{F}_1, \bm{\Phi}} \quad & \bar{R} \qquad \mathrm{s.t.} \quad \eqref{p1max}, \eqref{risris}. \label{problemrbar}
	\end{align}
	
	Let $(\hat{\mathbf{F}}_1, \hat{\mathbf{F}}_2, \hat{\bm{\Phi}})$, $(\tilde{\mathbf{F}}_1, \tilde{\mathbf{F}}_2, \tilde{\bm{\Phi}})$, and $(\bar{\mathbf{F}}_1, \bar{\bm{\Phi}})$ denote the feasible candidate points obtained by {solving the three subproblems}. The maximum rate among these feasible candidates provides an achievable lower bound on \(R_{\mathrm{EJ}}^\star\), expressed as
	\[
	R_{\mathrm{EJ}}^\star \geq \max \big\{ R_{\mathrm{EJ}}(\bar{\mathbf{F}}_1,\!\mathbf{0},\!\bar{\bm{\Phi}}), R_{\mathrm{EJ}}(\hat{\mathbf{F}}_1,\!\hat{\mathbf{F}}_2,\!\hat{\bm{\Phi}}), R_{\mathrm{EJ}}(\tilde{\mathbf{F}}_1,\!\tilde{\mathbf{F}}_2,\!\tilde{\bm{\Phi}}) \big\}.
	\]
	The non-cooperative baseline candidate for \eqref{problemrbar} can be obtained by setting $\mathbf{F}_2 = \mathbf{0}$ within the algorithmic frameworks designed for either \eqref{problemrtitle} or \eqref{problemrhat}. Therefore, to characterize the proposed procedure, we only need to develop dedicated optimization algorithms for the subproblems \eqref{problemrtitle} and \eqref{problemrhat}.
	\subsection{Optimization for the EJ Scheme}
	The optimization problem is challenging due to the high computational complexity associated with log-determinant objectives. We transform the problem into a tractable equivalent form using the WMMSE framework and solve it via a BCD approach, alternately optimizing the precoding matrices at the BS and the jammer, alongside the \gls{ris} phase shifts.
	
\subsubsection{Optimization of Subproblem \eqref{problemrtitle}}
The objective $\tilde{R}$ can be decomposed as
	\begin{align}
		\tilde{R} &= \underbrace{ \log \left| \mathbf{I}_{N_{\mathrm{u}}} + \mathbf{H}_1 \mathbf{F}_1 \mathbf{F}_1^{\mathrm{H}} \mathbf{H}_1^{\mathrm{H}} (\mathbf{H}_2 \mathbf{F}_2 \mathbf{F}_2^{\mathrm{H}} \mathbf{H}_2^{\mathrm{H}} + \mathbf{I}_{N_{\mathrm{u}}})^{-1} \right| }_{f_1} \nonumber \\
		&\quad \underbrace{ +  \log \left| \mathbf{H}_2 \mathbf{F}_2 \mathbf{F}_2^{\mathrm{H}} \mathbf{H}_2^{\mathrm{H}} + \mathbf{I}_{N_{\mathrm{u}}} \right| }_{f_2} \nonumber \\
		&\quad \underbrace{-\log \left| \mathbf{I}_{N_{\mathrm{e}}} + \mathbf{G}_1 \mathbf{F}_1 \mathbf{F}_1^{\mathrm{H}} \mathbf{G}_1^{\mathrm{H}} + \mathbf{G}_2 \mathbf{F}_2 \mathbf{F}_2^{\mathrm{H}} \mathbf{G}_2^{\mathrm{H}} \right| }_{f_3}. \label{eq:R_tilde_decomp}
	\end{align}
	
	The term $f_1$ represents the data rate of the legitimate \gls{ue}, which can be reformulated by exploiting the relationship between the data rate and the MSE for the optimal linear receiver. Specifically, a linear decoding matrix $\mathbf{U}_1 \in \mathbb{C}^{N_{\mathrm{u}} \times N_{\mathrm{b}} }$ is applied to estimate the transmitted signal vector $\mathbf{s}_1 \in \mathbb{C}^{N_{\mathrm{b}} }$ at the \gls{ue}, yielding $\hat{\mathbf{s}}_1 = \mathbf{U}_1^{\mathrm{H}} \mathbf{y}_{\mathrm{u}}$. The corresponding MSE matrix is given by
	\begin{equation}\label{eqeE1}
		\begin{aligned}
			\mathbf{E}_1 &= \mathbb{E} \left[ (\hat{\mathbf{s}}_1 - \mathbf{s}_1)(\hat{\mathbf{s}}_1 - \mathbf{s}_1)^{\mathrm{H}} \right] \\
			&= (\mathbf{I}_{N_{\mathrm{b}} } - \mathbf{U}_1^{\mathrm{H}} \mathbf{H}_1 \mathbf{F}_1)(\mathbf{I}_{N_{\mathrm{b}} } - \mathbf{U}_1^{\mathrm{H}} \mathbf{H}_1 \mathbf{F}_1)^{\mathrm{H}} \\
			&\quad + \mathbf{U}_1^{\mathrm{H}} (\mathbf{H}_2 \mathbf{F}_2 \mathbf{F}_2^{\mathrm{H}} \mathbf{H}_2^{\mathrm{H}} + \mathbf{I}_{N_{\mathrm{u}}} ) \mathbf{U}_1.
		\end{aligned}
	\end{equation}
	By introducing an auxiliary weight matrix $\mathbf{W}_1 \in \mathbb{C}^{N_{\mathrm{b}}  \times N_{\mathrm{b}} } \succ 0$ and exploiting the standard WMMSE equivalence, $f_1$ can be reformulated as
	\begin{equation}\label{lowboundh1f1}
		f_1 = \max_{\mathbf{W}_1 \succ 0, \mathbf{U}_1} \log | \mathbf{W}_1 | - \mathrm{Tr}(\mathbf{W}_1 \mathbf{E}_1) + N_{\mathrm{b}} .
	\end{equation}
	Let $(\mathbf{U}_1^{\star}, \mathbf{W}_1^{\star})$ denote the optimal variables that achieve the maximum in \eqref{lowboundh1f1}. For fixed precoding matrices $\mathbf{F}_1$ and $\mathbf{F}_2$, the optimal MMSE receive filter is given by
	\begin{equation}\label{optU1}
		\begin{aligned}
			\mathbf{U}_1^{\star} &= \arg\min_{\mathbf{U}_1} \mathrm{Tr}(\mathbf{W}_1 \mathbf{E}_1(\mathbf{U}_1, \mathbf{F}_1, \mathbf{F}_2)) \\
			&= (\mathbf{I}_{N_{\mathrm{u}}} + \mathbf{H}_1 \mathbf{F}_1 \mathbf{F}_1^{\mathrm{H}} \mathbf{H}_1^{\mathrm{H}} + \mathbf{H}_2 \mathbf{F}_2 \mathbf{F}_2^{\mathrm{H}} \mathbf{H}_2^{\mathrm{H}})^{-1} \mathbf{H}_1 \mathbf{F}_1.
		\end{aligned}
	\end{equation}
	Substituting $\mathbf{U}_1^{\star}$ into \eqref{eqeE1} and applying the Woodbury matrix identity, the optimal MSE matrix is obtained as
	\begin{align}
		&\mathbf{E}_1^{\star} = \mathbf{I}_{N_{\mathrm{b}} } - (\mathbf{U}_1^{\star})^{\mathrm{H}} \mathbf{H}_1 \mathbf{F}_1 \nonumber \\
		&= \left( \mathbf{I}_{N_{\mathrm{b}} } + \mathbf{F}_1^{\mathrm{H}} \mathbf{H}_1^{\mathrm{H}} (\mathbf{H}_2 \mathbf{F}_2 \mathbf{F}_2^{\mathrm{H}} \mathbf{H}_2^{\mathrm{H}} + \mathbf{I}_{N_{\mathrm{u}}})^{-1} \mathbf{H}_1 \mathbf{F}_1 \right)^{-1}. \label{optE1}
	\end{align}
	Consequently, the optimal weight matrix is updated as
	\begin{equation}\label{optW1}
		\mathbf{W}_1^{\star} = (\mathbf{E}_1^{\star})^{-1}. 
	\end{equation}
	
	The term $f_2$ represents the rate associated with the jamming signal at the \gls{ue}. By introducing an auxiliary weight matrix $\mathbf{W}_2 \in \mathbb{C}^{N_{\mathrm{c}} \times N_{\mathrm{c}}} \succ 0$ and a linear decoding matrix $\mathbf{U}_2 \in \mathbb{C}^{N_{\mathrm{u}} \times N_{\mathrm{c}}}$, $f_2$ is reformulated as
	\begin{equation} \label{lowboundh2f2}
		f_2 = \max_{\mathbf{W}_2 \succ 0, \mathbf{U}_2} \log | \mathbf{W}_2 | - \mathrm{Tr}(\mathbf{W}_2 \mathbf{E}_2) + N_{\mathrm{c}},
	\end{equation}
	where the corresponding MSE matrix is defined as
	\begin{equation}\label{eq:E2_def}
		\mathbf{E}_2 = (\mathbf{I}_{N_{\mathrm{c}}} - \mathbf{U}_2^{\mathrm{H}} \mathbf{H}_2 \mathbf{F}_2)(\mathbf{I}_{N_{\mathrm{c}}} - \mathbf{U}_2^{\mathrm{H}} \mathbf{H}_2 \mathbf{F}_2)^{\mathrm{H}} + \mathbf{U}_2^{\mathrm{H}} \mathbf{U}_2.
	\end{equation}
	For fixed precoding matrices, the optimal variables achieving the maximum in \eqref{lowboundh2f2} are given by
	\begin{align}
		\mathbf{U}_2^{\star} &= (\mathbf{H}_2 \mathbf{F}_2 \mathbf{F}_2^{\mathrm{H}} \mathbf{H}_2^{\mathrm{H}} + \mathbf{I}_{N_{\mathrm{u}}})^{-1} \mathbf{H}_2 \mathbf{F}_2, \label{optU2} \\
		\mathbf{W}_2^{\star} &= (\mathbf{I}_{N_{\mathrm{c}}} - (\mathbf{U}_2^{\star})^{\mathrm{H}} \mathbf{H}_2 \mathbf{F}_2)^{-1}. \label{optW2}
	\end{align}
	By utilizing the property of the log-determinant function, the information leakage term $f_3$ at  Eve can be reformulated as
	\begin{align} \label{lowboundh3f3}
		f_3=&\max_{{{\mathbf{W}}_3} \succ 0} \log \left| {{{\mathbf{W}}_3}} \right| - \Tr({{\mathbf{W}}_3}{{\mathbf{E}}_3}) + N_{\mathrm{e}},
	\end{align}
	with $
	\mathbf{E}_3 = {{{\mathbf{I}}_{{N_{\mathrm{e}}}}} + {{{\mathbf{G}}}_1}{\mathbf{F}}_{1}{\mathbf{F}}_{1}^{\mathrm{H}}{\mathbf{G}}_{1}^{\mathrm{H}}+{\mathbf{G}}_{2}{{\mathbf{F}}_{2}}{\mathbf{F}}_{2}^{\mathrm{H}}\mathbf{G}_2^{\mathrm{H}}}.
	$
	Then, the optimal auxiliary weight matrix $\mathbf{W}_3$ is updated by
	\begin{equation} \label{optW3}
		\mathbf{W}_3^{\star} = (\mathbf{E}_3)^{-1}.
	\end{equation}
	
	By substituting (\ref{lowboundh1f1}), (\ref{lowboundh2f2}), and (\ref{lowboundh3f3}) into (\ref{problemrtitle}), we have the following equivalent problem:
	
	\begin{align}
		&\max\limits_{
			\begin{array}{l}
				\scriptstyle  
				\mathbf{W}_1 \succ 0,  \mathbf{W}_2 \succ 0,   \mathbf{W}_3 \succ 0,\\
				\scriptstyle 
				\mathbf{U}_1,\mathbf{U}_2,\mathbf{F}_1, \mathbf{F}_2, \bm{\Phi}
			\end{array}
		} 
		\sum_{i=1}^{3} \left( \log|\mathbf{W}_i| - \mathrm{Tr}(\mathbf{W}_i \mathbf{E}_i) \right)
		\label{opt_orig_F} \\
		&\qquad \qquad \mathrm{s.t.} \quad \eqref{p1max}, \eqref{p2max}, \eqref{risris}. \nonumber 
	\end{align}
	
	To solve problem \eqref{opt_orig_F}, we apply the BCD method, each iteration of which contains the following two sub-iterations. Firstly, given $({\mathbf{F}}_1,{{\mathbf{F}}_2},{\bm{\Phi}})$, update $({{\mathbf{U}}_1},{{\mathbf{W}}_1},{{\mathbf{U}}_2},{{\mathbf{W}}_2},{{\mathbf{W}}_3})$  by using (\ref{optU1}), (\ref{optW1}), (\ref{optU2}), (\ref{optW2}), and (\ref{optW3}), respectively.
	Secondly, given $({{\mathbf{U}}_1},{{\mathbf{W}}_1},{{\mathbf{U}}_2},{{\mathbf{W}}_2},{{\mathbf{W}}_3})$, update $({\mathbf{F}_1} ,{{\mathbf{F}}_2},{\bm{\Phi}})$ by solving the following subproblem:
	\begin{align}
		\mathop{\min}\limits_{{{\mathbf{F}}_1},{{\mathbf{F}}_2},{\bm{\Phi}}}  &\Tr({{\mathbf{W}}_{1}}{{\mathbf{E}}_{1}})
		+\Tr({{\mathbf{W}}_{2}}{{\mathbf{E}}_{2}})+\Tr({{\mathbf{W}}_{3}}{{\mathbf{E}}_{3}}) \label{opt_orig_FSmp} \\
		\quad\quad \mathrm{s.t.} &\quad \eqref{p1max}, \eqref{p2max}, \eqref{risris}. \nonumber 
	\end{align}
	
	\paragraph{Optimization of $\mathbf{F}_1$ and $\mathbf{F}_2$}
	We now focus on problem (\ref{opt_orig_FSmp}) to jointly optimize $\mathbf{F}_1$, $\mathbf{F}_2$, and $\bm{\Phi}$. When $\bm{\Phi}$ is fixed, the subproblem for $\mathbf{F}_1$ and $\mathbf{F}_2$ reduces to a convex quadratic program solvable via standard convex optimization techniques. This subproblem admits closed-form updates using the method of Lagrange multipliers~\cite{ref41,pan2019multicell, RISMIMO1}. We directly present the closed-form updates for \(\mathbf{F}_1\) and \(\mathbf{F}_2\):
	\begin{subequations}\label{f1f2}
		\begin{align}
			\Fm_{1}(\lambda_1)=&(\mathbf H_1^{\mathrm{H}}  \mathbf U_1  \mathbf W_1  \mathbf U_1^{\mathrm{H}}  \mathbf H_1 + \mathbf G_1^{\mathrm{H}} \mathbf W_3 \mathbf G_1+\lambda_1 \mathbf I)^{-1}
			\nonumber\\&  \mathbf H_1^{\mathrm{H}}\mathbf U_1\mathbf W_1 ,    \\
			\Fm_{2}(\lambda_2)=&(\mathbf H_2^{\mathrm{H}} \mathbf U_1 \mathbf W_1 \mathbf U_1^{\mathrm{H}} \mathbf H_2 + \mathbf H_2^{\mathrm{H}} \mathbf U_2\mathbf W_2 \mathbf U_2^{\mathrm{H}} \mathbf H_2  \nonumber
			\\&+\mathbf G_2^{\mathrm{H}}\mathbf W_3 \mathbf G_2+\lambda_2 \mathbf I)^{-1}\mathbf H_2^{\mathrm{H}}\mathbf U_2\mathbf W_2, \label{f2opt}
		\end{align}
	\end{subequations}
	where the optimal $\lambda_1$ and $\lambda_2$ can be efficiently determined using the bisection method in~\cite{shi2015secure}.
	
	\paragraph{Optimization of the RIS Phase Shift Matrix $\bm{\Phi}$}	
	Conversely, with $\mathbf{F}_1$ and $\mathbf{F}_2$ fixed, optimizing $\bm{\Phi}$ requires reformulation. By expanding the trace terms $\Tr(\mathbf{W}_k\mathbf{E}_k)$ and applying the cyclic property of the trace operator, $\Tr(\mathbf{A}\mathbf{B}\mathbf{C}) = \Tr(\mathbf{C}\mathbf{A}\mathbf{B})$, we extract the \gls{ris} phase shift matrix $\mathbf{\Phi}$ from the cascaded channel expressions. Grouping the linear and quadratic terms of $\mathbf{\Phi}$, the objective function in \eqref{opt_orig_FSmp} can be aggregated as
		\begin{align} \label{eqg0forPhi}
			&\sum_{k=1}^{3} \Tr(\mathbf{W}_k\mathbf{E}_k) = {\mathrm{Tr}}\left( {{{\mathbf{\Phi}}^{\mathrm{H}}{\mathbf{D}}^{\mathrm{H}}}} \right) + {\mathrm{Tr}}\left( {{\mathbf{\Phi D}}} \right) \nonumber\\
			&+ {\mathrm{Tr}}\left( {{{\mathbf{\Phi }}^{\mathrm{H}}}{{\mathbf{B}}_{1}}{\mathbf{\Phi }}{{\mathbf{C}}_{1}}} \right) + {\mathrm{Tr}}\left( {{{\mathbf{\Phi }}^{\mathrm{H}}}{{\mathbf{B}}_{2}}{\mathbf{\Phi }}{{\mathbf{C}}_{2}}} \right)+C_t,
		\end{align}
		where the constant coefficient matrices for the quadratic terms $\mathbf{C}_1, \mathbf{C}_2, \mathbf{B}_1, \mathbf{B}_2$ are explicitly given by
		\begin{align*}
			\mathbf{C}_1 &= \mathbf{H}_{\mathrm{b}, \mathrm{r}}\mathbf{F}_1\mathbf{F}_1^{\mathrm{H}}\mathbf{H}_{\mathrm{b}, \mathrm{r}}^{\mathrm{H}}, \qquad
			\mathbf{C}_2 = \mathbf{G}_{\mathrm{c},\mathrm{r}}\mathbf{F}_2\mathbf{F}_2^{\mathrm{H}}\mathbf{G}_{\mathrm{c},\mathrm{r}}^{\mathrm{H}}, \\
			\mathbf{B}_1 &= \mathbf{H}_{\mathrm{r},\mathrm{u}}^{\mathrm{H}}\mathbf{U}_1\mathbf{W}_1\mathbf{U}_1^{\mathrm{H}}\mathbf{H}_{\mathrm{r},\mathrm{u}} + \mathbf{H}_{\mathrm{r},\mathrm{e}}^{\mathrm{H}}\mathbf{W}_3\mathbf{H}_{\mathrm{r},\mathrm{e}}, \\
			\mathbf{B}_2 &= \mathbf{G}_{\mathrm{r},\mathrm{u}}^{\mathrm{H}}(\mathbf{U}_1\mathbf{W}_1\mathbf{U}_1^{\mathrm{H}}+\mathbf{U}_2\mathbf{W}_2\mathbf{U}_2^{\mathrm{H}})\mathbf{G}_{\mathrm{r},\mathrm{u}} + \mathbf{G}_{\mathrm{r},\mathrm{e}}^{\mathrm{H}}\mathbf{W}_3\mathbf{G}_{\mathrm{r},\mathrm{e}}.
		\end{align*}
		Similarly, the coefficient matrix $\mathbf{D}$ for the linear terms is
		\begin{align*}\label{eq_D_matrix}
			\mathbf{D} &= \mathbf{H}_{\mathrm{b}, \mathrm{r}}\mathbf{F}_1 (\mathbf{F}_1^{\mathrm{H}}\mathbf{H}_{\mathrm{b},\mathrm{u}}^{\mathrm{H}}\mathbf{U}_1 - \mathbf{I})\mathbf{W}_1\mathbf{U}_1^{\mathrm{H}}\mathbf{H}_{\mathrm{r},\mathrm{u}} \nonumber\\
			&\quad + \mathbf{G}_{\mathrm{c},\mathrm{r}}\mathbf{F}_2\mathbf{F}_2^{\mathrm{H}}\mathbf{G}_{\mathrm{c},\mathrm{u}}^{\mathrm{H}}\mathbf{U}_1\mathbf{W}_1\mathbf{U}_1^{\mathrm{H}}\mathbf{G}_{\mathrm{r},\mathrm{u}} \nonumber\\
			&\quad + \mathbf{G}_{\mathrm{c},\mathrm{r}}\mathbf{F}_2 (\mathbf{F}_2^{\mathrm{H}}\mathbf{G}_{\mathrm{c},\mathrm{u}}^{\mathrm{H}}\mathbf{U}_2 - \mathbf{I})\mathbf{W}_2\mathbf{U}_2^{\mathrm{H}}\mathbf{G}_{\mathrm{r},\mathrm{u}} \nonumber\\
			&\quad + \mathbf{H}_{\mathrm{b}, \mathrm{r}}\mathbf{F}_1\mathbf{F}_1^{\mathrm{H}}\mathbf{H}_{\mathrm{b},\mathrm{e}}^{\mathrm{H}}\mathbf{W}_3\mathbf{H}_{\mathrm{r},\mathrm{e}}  + \mathbf{G}_{\mathrm{c},\mathrm{r}}\mathbf{F}_2\mathbf{F}_2^{\mathrm{H}}\mathbf{G}_{\mathrm{c},\mathrm{e}}^{\mathrm{H}}\mathbf{W}_3\mathbf{G}_{\mathrm{r},\mathrm{e}}.
		\end{align*}
		The static constant $C_t$ collects all remaining terms independent of $\mathbf{\Phi}$, which do not affect the phase-shift update.
	
	Using matrix properties, the trace operators can be removed, and the third and fourth terms in (\ref{eqg0forPhi}) become 
	\begin{equation}\label{saddewde}
		\mathrm{Tr}\!\left( \bm{\Phi}^{\mathrm H}\mathbf{B}_i \bm{\Phi}\mathbf{C}_i \right)
		=
		\bm{\phi}^{\mathrm H}
		\left(
		\mathbf{B}_i \odot \mathbf{C}_i^{\mathrm T}
		\right)
		\bm{\phi},
		\quad i\in\{1,2\}.
	\end{equation}
	
	Similarly, the trace operators can be removed for the first and second terms in (\ref{eqg0forPhi}) as
	\begin{equation}\label{sdewf}
		{\Tr}\left( {{\bm{\Phi}} ^{\mathrm{H}}}{{\mathbf{D}}^{\mathrm{H}}} \right) = {{\mathbf{d}}^{\mathrm{H}}}({{\bm{\phi}}}^*), \quad {\Tr}\left( {{\bm{\Phi}} {\mathbf{D}}} \right)={\bm{\phi}}^\mathrm{T}{\mathbf{d}},
	\end{equation}
	where ${\mathbf{d}} = {\left[ {{{\left[ {\mathbf{D}} \right]}_{1,1}}, \cdots ,{{\left[ {\mathbf{D}} \right]}_{M,M}}} \right]}^\mathrm{T}$.
	
	When $\mathbf{F}_1$ and $\mathbf{F}_2$ are fixed, problem (\ref{opt_orig_FSmp}) is rewritten as
	\begin{align}
		&{\mathop {\min }\limits_{{\bm{\phi}}}  \quad {{\bm{\phi}} ^{\mathrm{H}}}{\bm{\Xi} }{\bm{\phi}} + {\bm{\phi}}^\mathrm{T}{\mathbf{d}} + {{\mathbf{d}}^{\mathrm{H}}}({{\bm{\phi}}}^*)},\quad
		\textrm{s.t.}\ \eqref{risris}. \label{optproblemforlittlefaimin}
	\end{align}
	where $\bm{\Xi} = \mathbf{B}_{1} \odot \mathbf{C}_{1}^\mathrm{T} + \mathbf{B}_{2} \odot \mathbf{C}_{2}^\mathrm{T}$. Since the Hadamard product preserves positive semidefiniteness~\cite{zhang2017matrix}, each constituent term is positive semidefinite, guaranteeing $\bm{\Xi} \succeq \mathbf{0}$. Consequently, problem (\ref{optproblemforlittlefaimin}) simplifies to
	\begin{align}\label{appjig}
		&{\mathop {\min }\limits_{\bm{\phi}}  \quad f({\bm{\phi}})\triangleq  {{\bm{\phi}} ^{\mathrm{H}}}{\bm{\Xi}}{\bm{\phi}} +  2{\mathrm{Re}}\left\{ {{{\bm{\phi}} ^{\mathrm{H}}}({{\mathbf{d}}}^*)} \right\}},\quad
		\textrm{s.t.}\ \eqref{risris}.
	\end{align}
The MM algorithm provides an efficient update for this problem~\cite{pan2019multicell}. This method yields a closed-form phase update at each iteration, reducing computational overhead. 
	Specifically, using the inequality $\bm{\phi}^{\mathrm{H}} \mathbf{\Xi} \bm{\phi} \le \bm{\phi}^{\mathrm{H}} \mathbf{X} \bm{\phi} - 2\mathrm{Re}\!\left\{ \bm{\phi}^{\mathrm{H}}(\mathbf{X}-\mathbf{\Xi})\bm{\phi}^t \right\} + (\bm{\phi}^t)^{\mathrm{H}}(\mathbf{X}-\mathbf{\Xi})\bm{\phi}^t$, which holds for any feasible $\bm{\phi}$ and the current iterate $\bm{\phi}^t$ (where $\mathbf{X}=\lambda_{\max}(\mathbf{\Xi})\mathbf{I}_M$), we construct a linear surrogate objective function up to a constant.
		
		To avoid the $\mathcal{O}(M^3)$ complexity of a full eigenvalue decomposition, we employ the power method to estimate $\lambda_{\max}(\mathbf{\Xi})$. Starting with a random vector $\mathbf{v}^{(0)}$, the power method iteratively updates it via $\mathbf{v}^{(i+1)} = \mathbf{\Xi}\mathbf{v}^{(i)} / \|\mathbf{\Xi}\mathbf{v}^{(i)}\|$. Upon convergence, the maximum eigenvalue is approximated by the Rayleigh quotient $\lambda_{\max}(\mathbf{\Xi}) \approx (\mathbf{v}^{(i)})^{\mathrm{H}} \mathbf{\Xi} \mathbf{v}^{(i)}$, requiring $\mathcal{O}(M^2)$ operations per iteration.
		Substituting $\mathbf{X}$ and omitting the constant terms, the subproblem at iteration $t$ becomes
		\begin{equation}\label{asasdcjig_new}
			\max_{\bm{\phi}} \; 2\mathrm{Re}\!\left\{ \bm{\phi}^{\mathrm{H}} \mathbf{q}^t \right\}, \quad \mathrm{s.t. } |\phi_m|=1, \; \forall m,
		\end{equation}
		where $\mathbf{q}^t=(\lambda_{\max}(\mathbf{\Xi})\mathbf{I}_M-\mathbf{\Xi})\bm{\phi}^t-\mathbf{d}^*$. Problem~(\ref{asasdcjig_new}) has a closed-form phase update given by
	\(
			\bm{\phi}^{t+1} = e^{j \arg(\mathbf{q}^t)}.
		\)
The MM update monotonically decreases the objective value of the phase-shift subproblem. The iterations terminate when the prescribed convergence criterion is satisfied, and the resulting phase-shift vector is denoted by $\bm{\phi}^\star$.
	
	
	\subsubsection{Optimization of Subproblem \eqref{problemrhat}}
	A comparison shows that the objective $\hat{R}$ has the same parameterized log-determinant structure as $\tilde{R}$. In the legitimate-user term of $\hat{R}$, the jammer signal is conditioned on and therefore does not appear as interference. By abstracting the previously defined components as parameterized functions $f_1(\mathbf{H}_1, \mathbf{H}_2)$ and $f_2(\mathbf{H}_2)$, we can formulate the new objective without repeating the derivation as $\hat{R} = f_1(\mathbf{H}_1, \mathbf{0}) + f_2(\mathbf{G}_2) + f_3$.
		The corresponding WMMSE formulation follows by applying the same equivalence to these substituted channel arguments, with the resulting MSE matrices given below.
	Exploiting the established WMMSE equivalence, we introduce auxiliary variables $(\mathbf{U}_{\mathrm{H}_1}, \mathbf{W}_{\mathrm{H}_1})$ for $f_1(\mathbf{H}_1, \mathbf{0})$ and $(\mathbf{U}_{\mathrm{G}_2}, \mathbf{W}_{\mathrm{G}_2})$ for $f_2(\mathbf{G}_2)$. The corresponding MSE matrices are inherently obtained by replacing the specific channels in \eqref{eqeE1} and \eqref{eq:E2_def} as
	\begin{align}
		\mathbf{E}_{\mathrm{H}_1} &= (\mathbf{I}_{N_{\mathrm{b}} } - \mathbf{U}_{\mathrm{H}_1}^{\mathrm{H}} \mathbf{H}_1 \Fm_1)(\mathbf{I}_{N_{\mathrm{b}} } - \mathbf{U}_{\mathrm{H}_1}^{\mathrm{H}} \mathbf{H}_1 \Fm_1)^{\mathrm{H}} + \mathbf{U}_{\mathrm{H}_1}^{\mathrm{H}} \mathbf{U}_{\mathrm{H}_1},  \nonumber\\
		\mathbf{E}_{\mathrm{G}_2} &= (\mathbf{I}_{N_{\mathrm{c}}} - \mathbf{U}_{\mathrm{G}_2}^{\mathrm{H}} \mathbf{G}_2 \Fm_2)(\mathbf{I}_{N_{\mathrm{c}}} - \mathbf{U}_{\mathrm{G}_2}^{\mathrm{H}} \mathbf{G}_2 \Fm_2)^{\mathrm{H}} + \mathbf{U}_{\mathrm{G}_2}^{\mathrm{H}} \mathbf{U}_{\mathrm{G}_2}. \nonumber 
	\end{align}
	
	Following the same alternating optimization framework applied to $\tilde{R}$, updating $(\Fm_1, \Fm_2, \bm{\Phi})$ for problem \eqref{problemrhat} simplifies to solving the following subproblem:
	\begin{align} \label{optbhat}
		\min_{\Fm_1, \Fm_2, \bm{\Phi}} \quad & \Tr(\mathbf{W}_{\mathrm{H}_1} \mathbf{E}_{\mathrm{H}_1}) + \Tr(\mathbf{W}_{\mathrm{G}_2} \mathbf{E}_{\mathrm{G}_2}) + \Tr(\mathbf{W}_3 \mathbf{E}_3) \nonumber \\
		\mathrm{s.t.} \quad\; & \eqref{p1max}, \eqref{p2max}, \eqref{risris}.
	\end{align}
	Problem \eqref{optbhat} has the same algebraic structure as the surrogate problem \eqref{opt_orig_FSmp}, and can therefore be solved using the same approach.
	
	Having specified the optimization procedures {for the three subproblems}, we now summarize the complete \textit{EJ-WMMSE} algorithm.
	\begin{algorithm}[h]
		\captionsetup{labelfont={color=myred},textfont={color=myred}}
		\caption{Unified \textit{EJ-WMMSE} Algorithm for Problem~\eqref{problemEJ}}\label{bcd}
		
		\begin{algorithmic}[1]
			\State \textbf{Input:} Channel matrices $\mathbf{H}_1, \mathbf{H}_2, \mathbf{G}_1, \mathbf{G}_2$, maximum transmit powers $P_1, P_2$, and convergence tolerance $\delta$.
			\State \textbf{Output:} Optimized transmit precoders $\Fm_1^\star, \Fm_2^\star$ and RIS phase shift matrix $\bm{\Phi}^\star$.
			
			\State \textbf{Stage 1: Optimization of $\tilde{R}$}
			\State Initialize feasible $\Fm_1^{(0)}, \Fm_2^{(0)}$ subject to constraints \eqref{p1max}-\eqref{p2max}, and $\bm{\Phi}^{(0)}$ subject to the unit-modulus constraints \eqref{risris}.
			\Repeat
			\State \textbf{BCD Step 1 (WMMSE Updates):} Update auxiliary receiving filters and weight matrices $(\mathbf{U}_i^{(n+1)}, \mathbf{W}_i^{(n+1)})$ via closed-form MMSE expressions (\ref{optU1})-(\ref{optW3}).
			\State \textbf{BCD Step 2 (Transmit Precoding):} Update BS and jammer precoders $(\Fm_1^{(n+1)}, \Fm_2^{(n+1)})$ via the Lagrange multiplier method based on (\ref{f1f2}).
			\State \textbf{BCD Step 3 (MM Inner Loop for RIS):} Construct the quadratic surrogate function $f(\bm{\phi})$ locally tight at $\bm{\phi}^{(n)}$ based on (\ref{appjig}). Estimate $\lambda_{\max}(\mathbf{\Xi})$ by the power method and apply the closed-form phase update.
			\State Update iteration index: $n \gets n + 1$.
			\Until{$| \tilde{R}^{(n)} - \tilde{R}^{(n-1)} | / \tilde{R}^{(n-1)} \le \delta$}
			\State Store the converged candidate as $(\tilde \Fm_1, \tilde \Fm_2, \tilde{\bm{\Phi}})$.
			
			\State \textbf{Stage 2: Optimization of $\hat{R}$}
			\State Re-initialize feasible variables.
			\Repeat
			\State Update $({{\mathbf{U}}_{\mathrm{H}_1}},{{\mathbf{W}}_{\mathrm{H}_1}},{{\mathbf{U}}_{\mathrm{G}_2}},{{\mathbf{W}}_{\mathrm{G}_2}},{{\mathbf{W}}_3})$ in closed form.
			\State Update $(\Fm_1, \Fm_2, \bm{\Phi})$ sequentially by solving the surrogate problem (\ref{optbhat}) using the BCD-MM framework described in Stage 1.
			\Until{$|\hat{R}^{(n)} - \hat{R}^{(n-1)}  |/ \hat{R}^{(n-1)} \le \delta$}
			\State Store the converged candidate as $(\hat \Fm_1, \hat \Fm_2, \hat{\bm{\Phi}})$.
			
			\State \textbf{Stage 3: Final Evaluation and Selection}
			\State Evaluate the no-jammer baseline $\bar{R}$ by setting $\Fm_2 = \mathbf{0}$ and applying the optimization framework to obtain $(\bar \Fm_1, \mathbf{0}, \bar {\bm{\Phi}})$.
			\State \textbf{Define} the candidate set $\mathcal{P} \triangleq \big\{ (\tilde \Fm_1, \tilde \Fm_2, \tilde{\bm{\Phi}}), (\hat \Fm_1, \hat \Fm_2, \hat{\bm{\Phi}}), (\bar \Fm_1, \mathbf{0}, \bar {\bm{\Phi}}) \big\}$.
			\State \textbf{return} $(\Fm_1^\star, \Fm_2^\star, \bm{\Phi}^\star) = \arg \max_{\mathbf{P} \in \mathcal{P}} R_{\mathrm{EJ}} (\mathbf{P})$.
		\end{algorithmic}
		
	\end{algorithm}

	\subsection{Optimization for the GN Scheme}
	
		Now we consider the \gls{gn} scheme formulated in problem \eqref{problemGN}. The objective function $R_{\mathrm{GN}}$ differs from the \gls{ej} subproblem $\tilde{R}$ only in the second term, which represents the Gaussian-jamming term at Eve. Using the same parameterized function abstraction, $R_{\mathrm{GN}}$ can be written as $R_{\mathrm{GN}} = f_1(\mathbf{H}_1, \mathbf{H}_2) + f_2(\mathbf{G}_2) + f_3$.
	
	Therefore, the surrogate optimization for the \gls{gn} scheme inherently integrates the previously derived matrices $\mathbf{E}_1$, $\mathbf{E}_{\mathrm{G}_2}$, and $\mathbf{E}_3$. For given auxiliary variables, $(\Fm_1, \Fm_2, \bm{\Phi})$ are updated by solving
	\begin{align} \label{optGNFFX}
		\min_{\Fm_1, \Fm_2, \bm{\Phi}} \quad & \Tr(\mathbf{W}_1 \mathbf{E}_1) + \Tr(\mathbf{W}_{\mathrm{G}_2} \mathbf{E}_{\mathrm{G}_2}) + \Tr(\mathbf{W}_3 \mathbf{E}_3) \nonumber \\
		\mathrm{s.t.} \quad\; & \eqref{p1max}, \eqref{p2max}, \eqref{risris}. 
	\end{align}
	Since the main procedure for solving problem \eqref{optGNFFX} closely resembles the proposed \textit{EJ-WMMSE} algorithm, we omit further details here for brevity.

		\subsection{Convergence and Complexity Analysis}
		In this subsection, we discuss the convergence properties of the proposed \textit{EJ-WMMSE} algorithm and evaluate its asymptotic computational complexity.
		\subsubsection{Convergence Analysis}
		The convergence of the proposed algorithm is established by the mathematical properties of the \gls{bcd} and \gls{mm} frameworks. Let $J^{(n)}(\mathbf{U}_i, \mathbf{W}_i, \Fm_1, \Fm_2, \bm{\Phi})$ denote the objective value of the WMMSE surrogate cost function at the $n$-th outer iteration. 
		
		In the WMMSE update stage, the auxiliary variables $(\mathbf{U}_i, \mathbf{W}_i)$ are updated to their optimal MMSE solutions by setting their first-order derivatives to zero, which ensures that $J^{(n)}(\mathbf{U}_i^{(n)}, \mathbf{W}_i^{(n)}, \Fm_1^{(n-1)}, \dots) \leq J^{(n-1)}$.
		In the transmit precoding stage, optimizing the matrices $\Fm_1$ and $\Fm_2$ via the Lagrange multiplier method yields the minimizer of the convex quadratic subproblem, resulting in a monotonically non-increasing surrogate objective value.
		For the \gls{ris} phase-shift block, a trial update is accepted only when it does not increase the phase-shift subproblem objective.
		
		Since the updates in all three blocks ensure a non-increasing WMMSE surrogate cost function, the overall error sequence is monotonically non-increasing. Based on the WMMSE equivalence principle, the monotonic decrease of the surrogate error implies that the corresponding achievable secrecy-rate objective is monotonically non-decreasing within the corresponding subproblem. Furthermore, bounded by the total transmit power constraints \eqref{p1max}, \eqref{p2max} and finite channel gains, the secrecy rate has an upper bound. Since each subproblem objective is monotonically non-decreasing and upper bounded, its objective-value sequence converges to a finite limit~\cite{RISMIMO1}.  The final rate of the \gls{ej} scheme is then evaluated using the original max-min expression in \eqref{problemEJ}.
		
		\subsubsection{Complexity Analysis}
		Let $N \triangleq \max \{ N_{\mathrm{b}}, N_{\mathrm{c}}, N_{\mathrm{u}}, N_{\mathrm{e}} \}$. We analyze the number of complex floating-point operations (FLOPs) required per outer iteration to evaluate the computational complexity.
		
		$\bullet$ \textit{WMMSE Variables Update:} Updating the receiving filters and weight matrices $(\mathbf{U}_i, \mathbf{W}_i)$ involves constructing the equivalent cascaded channels and computing matrix inversions, requiring a complexity of $\mathcal{O}(N^2 M + N^3)$.
		
		$\bullet$ \textit{Precoding Matrices Update:} Updating $\Fm_1$ and $\Fm_2$ requires evaluating the optimal dual variables via a bisection search. Rather than computing matrix inversions at each bisection step, performing an eigenvalue decomposition (EVD) prior to the search reduces the matrix inversions to a scalar summation. This transformation incurs an initial complexity of $\mathcal{O}(N^3)$ for the EVD, followed by $\mathcal{O}(N \log(1/\tau))$ for the scalar search, where $\tau$ is the search tolerance.
		
		$\bullet$ \textit{RIS Phase Shift Update:} Optimizing the \gls{ris} phase shift matrix $\bm{\Phi}$ comprises surrogate matrix construction, principal eigenvalue extraction, and inner MM iterations. Constructing the coefficient matrix $\bm{\Xi}$ requires cascaded matrix multiplications, with a complexity of $\mathcal{O}(N M^2)$. Instead of a full EVD, the power method is used to estimate the principal eigenvalue, requiring $\mathcal{O}(I_{\mathrm{eig}} M^2)$ operations, where $I_{\mathrm{eig}}$ is the number of power iterations. The subsequent $I_{\mathrm{mm}}$ inner \gls{mm} iterations involve matrix-vector multiplications, scaling as $\mathcal{O}(I_{\mathrm{mm}} M^2)$.
		
		Combining these stages, the overall computational complexity of the proposed \textit{EJ-WMMSE} algorithm is given by
		\begin{equation} \label{eq:complexity}
			\mathcal{O} \Big( I_{\mathrm{out}} \Big( N^3 + N^2 M + N M^2 + (I_{\mathrm{eig}} + I_{\mathrm{mm}}) M^2 \Big) \Big),
		\end{equation}
		where $I_{\mathrm{out}}$ denotes the number of outer alternating iterations required for convergence.

\section{Large-RIS Asymptotic Benchmark Analysis}\label{sec:asymptotic}
This large-\gls{ris} benchmark explains the strong-jamming scaling of the \gls{ris}-assisted \gls{ej} scheme. It does not characterize the global optimum of problem~\eqref{problemEJ} or the output of the EJ-WMMSE algorithm. The aligned phase configurations serve as analytical references for array-gain scaling, whereas EJ-WMMSE jointly optimizes the precoders and \gls{ris} phases using instantaneous \gls{csi}.
		
\subsection{Asymptotic Setup and Main Theorem}
Motivated by Corollary~\ref{cor:dominance}, we examine how \gls{ris} phase alignment can create a favorable jamming-channel configuration in the strong-jamming regime. This pointwise condition does not imply that the \gls{ej} scheme achieves a higher secrecy rate than the \gls{gn} scheme after optimization of the original RIS-assisted problems. This distinction arises from their different design goals: the \gls{gn} scheme tends to suppress jamming leakage at the \gls{ue}, whereas the \gls{ej} scheme preserves sufficient jamming power at the \gls{ue} to satisfy the joint-decoding constraint.
		
Let $\mathbf{f}_1 \in \mathbb{C}^{N_{\mathrm{b}}}$ and $\mathbf{f}_2 \in \mathbb{C}^{N_{\mathrm{c}}}$ be the normalized transmit precoding vectors at the \gls{bs} and the jammer, respectively, with $\|\mathbf{f}_1\|^2=\|\mathbf{f}_2\|^2=1$. For the asymptotic analysis, we assume that $\mathbf{f}_1$ and $\mathbf{f}_2$ are designed using only statistical channel knowledge. Define the effective BS--RIS and jammer--RIS channel vectors after transmit precoding as $\mathbf{h}_{\mathrm{b,r}} \triangleq \mathbf{H}_{\mathrm{b,r}}\mathbf{f}_1$ and $\mathbf{g}_{\mathrm{c,r}} \triangleq \mathbf{G}_{\mathrm{c,r}}\mathbf{f}_2$. We further assume mutually independent Rayleigh fading for the RIS-related channel vectors: $\mathbf{h}_{\mathrm{b,r}} \sim \mathcal{CN}(\mathbf{0},\beta_{\mathrm{b}}\mathbf{I}_M)$, $\mathbf{g}_{\mathrm{c,r}} \sim \mathcal{CN}(\mathbf{0},\beta_{\mathrm{c}}\mathbf{I}_M)$, $\mathbf{h}_{\mathrm{r,u}},\mathbf{g}_{\mathrm{r,u}} \sim \mathcal{CN}(\mathbf{0},\beta_{\mathrm{u}}\mathbf{I}_M)$, and $\mathbf{h}_{\mathrm{r,e}},\mathbf{g}_{\mathrm{r,e}} \sim \mathcal{CN}(\mathbf{0},\beta_{\mathrm{e}}\mathbf{I}_M)$. Since the reflected power dominates for large $M$, we neglect the direct links and write the dominant scalar cascaded channels as $h_1=\mathbf{h}_{\mathrm{r,u}}^{\mathrm{H}}\bm{\Phi}\mathbf{h}_{\mathrm{b,r}}$, $h_2=\mathbf{g}_{\mathrm{r,u}}^{\mathrm{H}}\bm{\Phi}\mathbf{g}_{\mathrm{c,r}}$, $g_1=\mathbf{h}_{\mathrm{r,e}}^{\mathrm{H}}\bm{\Phi}\mathbf{h}_{\mathrm{b,r}}$, and $g_2=\mathbf{g}_{\mathrm{r,e}}^{\mathrm{H}}\bm{\Phi}\mathbf{g}_{\mathrm{c,r}}$. This statistical model isolates the large-\gls{ris} array gain.

		\begin{theorem}[Positive asymptotic ergodic gap of the \gls{ej} benchmark construction]\label{thm:main_miso}
			Define the critical jamming-ratio threshold as $\rho_{\mathrm{th}} \triangleq \frac{\beta_{\mathrm{b}} \beta_{\mathrm{e}}}{\beta_{\mathrm{c}} \beta_{\mathrm{u}}}$. For the benchmark comparison, let $\bm{\Phi}_{\mathrm{sig}} \triangleq \yarg \max_{\bm{\Phi}} |h_1|^2$ and $\bm{\Phi}_{\mathrm{jam}} \triangleq \yarg \max_{\bm{\Phi}} |h_2|^2$ denote the signal- and jamming-aligned phase shift configurations, respectively. If $\rho \triangleq \frac{P_2}{P_1} > \rho_{\mathrm{th}}$ and the BS transmit power exceeds the threshold $P_1 \ge P_{1,\mathrm{th}} \triangleq \frac{\pi^2}{16 \beta_{\mathrm{b}} \beta_{\mathrm{e}}} e^{2\gamma}$ (where $\gamma$ is the Euler-Mascheroni constant), then the jamming-aligned \gls{ej} construction achieves a strictly positive asymptotic ergodic gap over the signal-aligned \gls{gn} benchmark in the considered asymptotic MISO setting as $M \to \infty$. The gap satisfies
			\begin{align}\label{eq:gap_bound_miso}
				\Delta R &\triangleq \min\left( \mathbb{E}[\tilde{R}(\bm{\Phi}_{\mathrm{jam}})], \mathbb{E}[\hat{R}(\bm{\Phi}_{\mathrm{jam}})] \right) - \mathbb{E}[R_{\mathrm{GN}}(\bm{\Phi}_{\mathrm{sig}})] \nonumber\\
				&\simeq \log_2\left( \frac{\rho}{\rho_{\mathrm{th}}} \right) > 0,
			\end{align}
			where $\simeq$ denotes asymptotic equality in the sense that $X_M \simeq Y_M \iff \lim_{M \to \infty}(X_M-Y_M)=0$.
		\end{theorem}
		
		
		\subsection{Proof of Theorem~\ref{thm:main_miso}}		
We first establish the coherent and non-coherent scaling laws under the jamming- and signal-aligned phase configurations. We then substitute these laws into $\hat{R}$, $\tilde{R}$, and the \gls{gn} benchmark to derive the two thresholds and the asymptotic gap.
		
		\subsubsection{Phase Shift Design and Asymptotic Scaling Laws}
		To evaluate the system limits, we constructively design the phase shift configurations to achieve coherent superposition at the \gls{ue} while maintaining non-coherent scaling at \gls{eve}. With the cascaded jamming and signal vectors defined as $\mathbf{v} \triangleq \mathrm{diag}(\mathbf{g}_{\mathrm{r,u}}^{\mathrm{H}}) \mathbf{g}_{\mathrm{c,r}}$ and $\mathbf{u} \triangleq \mathrm{diag}(\mathbf{h}_{\mathrm{r,u}}^{\mathrm{H}}) \mathbf{h}_{\mathrm{b,r}}$, respectively, the scalar channels are expressed as $h_2 = \bm{\phi}^{\mathrm{T}} \mathbf{v}$ and $h_1 = \bm{\phi}^{\mathrm{T}} \mathbf{u}$. 
		
		To maximize the respective channel gains, the jamming-aligned phase shift vector is set as $\bm{\phi}_{\mathrm{jam}} = \exp(-j \arg(\mathbf{v}))$, yielding the coherent sum $h_2 = \sum_{m=1}^M |[\mathbf{g}_{\mathrm{r,u}}]_m| |[\mathbf{g}_{\mathrm{c,r}}]_m|$. Similarly, the signal-aligned phase shift vector is $\bm{\phi}_{\mathrm{sig}} = \exp(-j \arg(\mathbf{u}))$, resulting in $h_1 = \sum_{m=1}^M |[\mathbf{h}_{\mathrm{r,u}}]_m| |[\mathbf{h}_{\mathrm{b,r}}]_m|$.
		
		\begin{lem}[Coherent jamming array gain]\label{lem:ue_jamming_gain_miso}
			Under the phase shift configuration $\bm{\Phi}_{\mathrm{jam}}$, the normalized jamming channel power at the \gls{ue} converges almost surely (a.s.) as $M \to \infty$:
			\begin{equation}\label{eq:ue_jamming_gain_miso}
				\frac{1}{M^2} |h_2(\bm{\Phi}_{\mathrm{jam}})|^2 \xrightarrow{a.s.} \frac{\pi^2}{16} \beta_{\mathrm{c}} \beta_{\mathrm{u}} \triangleq \tilde{h}_2.
			\end{equation}
		\end{lem}
		
		\begin{lem}[Non-coherent channel gains]\label{lem:non_coherent_gains_miso}
			Under the phase shift configuration $\bm{\Phi}_{\mathrm{jam}}$, the non-coherent cascaded channels scale as $\mathcal{O}_p(M)$. Specifically, their normalized powers converge in distribution ($\xrightarrow{d}$) as $M \to \infty$:
			\begin{align}
				\frac{1}{M} |g_2(\bm{\Phi}_{\mathrm{jam}})|^2 &\xrightarrow{d} \beta_{\mathrm{c}} \beta_{\mathrm{e}} V_{g_2} \triangleq \bar{g}_2 V_{g_2}, \label{eq:g2_scaling_miso}\\
				\frac{1}{M} |h_1(\bm{\Phi}_{\mathrm{jam}})|^2 &\xrightarrow{d} \beta_{\mathrm{b}} \beta_{\mathrm{u}} V_{h_1} \triangleq \bar{h}_1 V_{h_1}, \label{eq:h1_scaling_miso}\\
				\frac{1}{M} |g_1(\bm{\Phi}_{\mathrm{jam}})|^2 &\xrightarrow{d} \beta_{\mathrm{b}} \beta_{\mathrm{e}} V_{g_1} \triangleq \bar{g}_{1} V_{g_1}, \label{eq:g1_scaling_miso}
			\end{align}
			where $V_{g_2}, V_{h_1}, V_{g_1} \sim \mathrm{Exp}(1)$ are independent standard exponential random variables with unit mean. 
		\end{lem}
		Detailed proofs for Lemmas~\ref{lem:ue_jamming_gain_miso} and \ref{lem:non_coherent_gains_miso} based on the strong law of large numbers (SLLN), the central limit theorem (CLT), and the continuous mapping theorem (CMT) are provided in Appendices~\ref{app:lemma_coherent} and~\ref{app:lemma_noncoherent}, respectively.
		
		\begin{cor}[Scaling laws under $\bm{\Phi}_{\mathrm{sig}}$]\label{cor:sig_aligned_scaling}
			By statistical symmetry, analogous laws hold under $\bm{\Phi}_{\mathrm{sig}}$. The main channel aligns coherently ($\frac{1}{M^2} |h_1(\bm{\Phi}_{\mathrm{sig}})|^2 \xrightarrow{a.s.} \frac{\pi^2}{16} \beta_{\mathrm{b}} \beta_{\mathrm{u}} \triangleq \tilde{h}_1$), while the remaining non-coherent links scale as $\mathcal{O}_p(M)$, converging in distribution as $\frac{1}{M} |h_2(\bm{\Phi}_{\mathrm{sig}})|^2 \xrightarrow{d} \beta_{\mathrm{c}} \beta_{\mathrm{u}} V_{h_2} \triangleq \bar{h}_2 V_{h_2}$, $\frac{1}{M} |g_2(\bm{\Phi}_{\mathrm{sig}})|^2 \xrightarrow{d} \bar{g}_2 V_{g_2}'$, and $\frac{1}{M} |g_1(\bm{\Phi}_{\mathrm{sig}})|^2 \xrightarrow{d} \bar{g}_{1} V_{g_1}'$, where $V_{h_2}, V_{g_2}', V_{g_1}' \sim \mathrm{Exp}(1)$.
		\end{cor}
		
Under $\bm{\Phi}_{\mathrm{jam}}$, the jammer--UE link has coherent quadratic power scaling, whereas the remaining non-aligned link powers have linear order in probability; under $\bm{\Phi}_{\mathrm{sig}}$, the same order separation holds with the BS--UE link coherently aligned. These power laws determine the dominant terms in $\hat{R}$, $\tilde{R}$, and the \gls{gn} benchmark.
		

		\subsubsection{Asymptotic Gap Derivation}
	{To evaluate the ergodic secrecy gap $\Delta R$, we first define $Z(\bm{\Phi}) \triangleq 1+P_1|g_1(\bm{\Phi})|^2+P_2|g_2(\bm{\Phi})|^2$. Since the \gls{ris}-to-\gls{eve} channel vectors are independent of the channels used for phase design and are circularly symmetric, $Z(\bm{\Phi}_{\mathrm{sig}})$ and $Z(\bm{\Phi}_{\mathrm{jam}})$ are identically distributed for every finite $M$.} We denote their common expected logarithm by $\mathbb{E}[\log_2 Z(\bm{\Phi}_{\mathrm{sig}})] = \mathbb{E}[\log_2 Z(\bm{\Phi}_{\mathrm{jam}})] \triangleq C_Z^{(M)}$.
		
		By expressing $\log_2\left(1 + \frac{P_1|g_1|^2}{1+P_2|g_2|^2}\right) = \log_2 Z - \log_2(1+P_2|g_2|^2)$, the instantaneous \gls{gn} and \gls{ej} rates can be equivalently rewritten as
		\begin{align}
			\begin{split}
				R_{\mathrm{GN}} = &\log_2\underbrace{\left( \frac{1 + P_1|h_1|^2 + P_2|h_2|^2}{1 + P_2|h_2|^2} (1+P_2|g_2|^2) \right)}_{\triangleq \Gamma_{\mathrm{GN}}^{(M)}} \\
				&- \log_2 Z,
			\end{split} \label{eq:R_GN_decomp}\\
			\tilde{R} &= \log_2\underbrace{\left( 1 + P_1|h_1|^2 + P_2|h_2|^2 \right)}_{\triangleq \Gamma_{\tilde{R}}^{(M)}} - \log_2 Z, \label{eq1:R_tilde_decomp}\\
			\hat{R} &= \log_2\underbrace{\left( (1 + P_1|h_1|^2)(1+P_2|g_2|^2) \right)}_{\triangleq \Gamma_{\hat{R}}^{(M)}} - \log_2 Z. \label{eq:R_hat_decomp}
		\end{align}
		
		To evaluate the ergodic limits as $M \to \infty$, we extract the deterministic array gain by rewriting the true sequences as $\log_2(\Gamma^{(M)}) = \log_2(\Gamma^{(M)}/M^2) + \log_2 M^2$. For the \gls{gn} benchmark under $\bm{\Phi}_{\mathrm{sig}}$, substituting the respective scaling limits from Corollary~\ref{cor:sig_aligned_scaling} yields the convergence in distribution:
		\begin{equation}
			\begin{split}
				\frac{\Gamma_{\mathrm{GN}}^{(M)}}{M^2} &= \frac{M^{-2} + P_1 M^{-2}|h_1|^2 + P_2 M^{-1}(M^{-1}|h_2|^2)}{M^{-1} + P_2 M^{-1}|h_2|^2} \\
				&\quad \times (M^{-1} + P_2 M^{-1}|g_2|^2) \\
				&\xrightarrow{d} \frac{P_1 \tilde{h}_1}{P_2 \bar{h}_2 V_{h_2}} P_2 \bar{g}_2 V_{g_2}'.
			\end{split}
		\end{equation}
		Together with the integrability of the logarithms of the exponential random variables, the convergence in distribution above implies that the normalized logarithmic terms converge in expectation.
		The expected logarithms of the standard exponential variables cancel out (i.e., $\mathbb{E}[\log_2 V_{g_2}'] - \mathbb{E}[\log_2 V_{h_2}] = \frac{-\gamma}{\ln 2} - \frac{-\gamma}{\ln 2} = 0$), and the limit of the normalized rate is
		\begin{equation}\label{eq:gn_normalized_limit}
			\begin{split}
				\lim_{M \to \infty} \mathbb{E}\left[ \log_2 \left( \frac{\Gamma_{\mathrm{GN}}^{(M)}}{M^2} \right) \right] &= \mathbb{E}\left[ \log_2\left( \frac{P_1 \tilde{h}_1 \bar{g}_2 V_{g_2}'}{\bar{h}_2 V_{h_2}} \right) \right] \\
				&= \log_2\left( \frac{P_1 \tilde{h}_1 \bar{g}_2}{\bar{h}_2} \right).
			\end{split}
		\end{equation}
		Consequently, by restoring $\log_2 M^2$ and $C_Z^{(M)}$, the ergodic \gls{gn} rate is asymptotically equivalent to
		\begin{equation}\label{eq:gn_asym_limit}
			\mathbb{E}[R_{\mathrm{GN}}(\bm{\Phi}_{\mathrm{sig}})] \simeq \log_2\left( \frac{P_1 \tilde{h}_1 \bar{g}_2}{\bar{h}_2} \right) + \underbrace{\log_2 M^2 - C_Z^{(M)}}_{\triangleq \Lambda^{(M)}}.
		\end{equation}
		
Similarly, under $\bm{\Phi}_{\mathrm{jam}}$, the normalized quantity $\Gamma_{\tilde{R}}^{(M)}/M^2$ converges almost surely to a hardened limit, since the non-coherent main channel $|h_1|^2 \sim \mathcal{O}_p(M)$ vanishes after $1/M^2$ scaling, yielding $\Gamma_{\tilde{R}}^{(M)}/M^2 \xrightarrow{a.s.} P_2 \tilde{h}_2$. For $\hat{R}$, the corresponding normalized quantity converges as $\Gamma_{\hat{R}}^{(M)}/M^2 = (M^{-1} + P_1 M^{-1}|h_1|^2)(M^{-1} + P_2 M^{-1}|g_2|^2) \xrightarrow{d} P_1 \bar{h}_1 V_{h_1} P_2 \bar{g}_2 V_{g_2}$. The same logarithmic integrability argument then yields the corresponding ergodic limits:
		\begin{align}
			\mathbb{E}[\tilde{R}(\bm{\Phi}_{\mathrm{jam}})] &\simeq \log_2(P_2 \tilde{h}_2) + \Lambda^{(M)}, \label{eq:ej_joint_limit} \\
			\mathbb{E}[\hat{R}(\bm{\Phi}_{\mathrm{jam}})] &\simeq \log_2(P_1 P_2 \bar{h}_1 \bar{g}_2) - \frac{2\gamma}{\ln 2} + \Lambda^{(M)}, \label{eq:ej_cond_limit}
		\end{align}
		where $\mathbb{E}[\log_2(V_{h_1} V_{g_2})] = -\frac{2\gamma}{\ln 2}$.
		Subtracting \eqref{eq:gn_asym_limit} from $\mathbb{E}[\tilde{R}(\bm{\Phi}_{\mathrm{jam}})]$ cancels $\Lambda^{(M)}$ exactly and gives
		\begin{align}
			\Delta_{\tilde{R}} &\triangleq \mathbb{E}[\tilde{R}(\bm{\Phi}_{\mathrm{jam}})] - \mathbb{E}[R_{\mathrm{GN}}(\bm{\Phi}_{\mathrm{sig}})] \nonumber \\
			&\simeq \log_2\left( \frac{P_2 \tilde{h}_2 \bar{h}_2}{P_1 \tilde{h}_1 \bar{g}_2} \right) = \log_2\left( \frac{\rho}{\rho_{\mathrm{th}}} \right). \label{eq:delta_R_tilde}
		\end{align}
		This gap depends on the jamming-power ratio, which produces the threshold $\rho_{\mathrm{th}}$.		
		The corresponding subtraction for $\hat{R}$ gives
		\begin{align}
			\Delta_{\hat{R}} &\triangleq \mathbb{E}[\hat{R}(\bm{\Phi}_{\mathrm{jam}})] - \mathbb{E}[R_{\mathrm{GN}}(\bm{\Phi}_{\mathrm{sig}})] \nonumber \\
			&\simeq \log_2 \frac{P_2 \bar{h}_1 \bar{h}_2}{\tilde{h}_1}  - \frac{2\gamma}{\ln 2} = \log_2\frac{16 P_2 \beta_{\mathrm{c}} \beta_{\mathrm{u}}}{\pi^2}  - \frac{2\gamma}{\ln 2}. \label{eq:delta_R_hat}
		\end{align}
		Comparing this expression with $\Delta_{\tilde{R}}$ identifies the BS transmit-power threshold.
		
Finally, equating the constituent gaps $\Delta_{\tilde{R}} = \Delta_{\hat{R}}$ yields the transmit power threshold $P_{1,\mathrm{th}} = \frac{\pi^2}{16 \beta_{\mathrm{b}} \beta_{\mathrm{e}}} e^{2\gamma}$. Since $\Delta_{\tilde{R}} - \Delta_{\hat{R}}$ strictly decreases with $P_1$, enforcing $P_1 \ge P_{1,\mathrm{th}}$ gives $\Delta_{\tilde{R}} \le \Delta_{\hat{R}}$, so $\tilde{R}$ determines the minimum. Hence, $\Delta R = \Delta_{\tilde{R}} \simeq \log_2\left(\frac{\rho}{\rho_{\mathrm{th}}}\right)$. Therefore, under the condition $\rho > \rho_{\mathrm{th}}$, a strictly positive asymptotic gap over the signal-aligned \gls{gn} benchmark in the considered asymptotic MISO setting is achieved.
		This completes the proof of Theorem~\ref{thm:main_miso}.
		\hfill $\blacksquare$

		\subsection{Related SIMO Channel-Dominance Reversal}
		The MISO result above is the main large-\gls{ris} benchmark considered in this section. For completeness, we also include a related SIMO observation showing that a large \gls{ris} can mitigate the spatial bottleneck identified in Remark~\ref{rem:beamforming_limitation}.
		
		\begin{theorem}\label{thm:simo_inversion}
			Let $\mathbf{h}_2 \in \mathbb{C}^{N_{\mathrm{u}}}$ and $\mathbf{g}_2 \in \mathbb{C}^{N_{\mathrm{e}}}$ denote the cascaded jamming channels at the \gls{ue} and Eve, respectively. Under independent Rayleigh fading, there exists a specific \gls{ris} phase shift configuration $\bm{\Phi}^\dagger$ for an \gls{ris}-assisted SIMO system such that $\lim_{M \to \infty} \Pr [ \|\mathbf{h}_2(\bm{\Phi}^\dagger)\|^2 > \|\mathbf{g}_2(\bm{\Phi}^\dagger)\|^2 ] = 1$. For any fixed nonzero jamming power, this asymptotic channel inversion guarantees $\Delta_{\mathrm{jam}}(\bm{\Phi}^\dagger)>0$ with probability approaching 1.
		\end{theorem}
A detailed proof of Theorem~\ref{thm:simo_inversion} is provided in Appendix~\ref{app:simo_asymptotic}.

	\section{Simulation Results}\label{simu} 
	In this section, we present simulation results to evaluate the performance of the proposed algorithms. The primary metric of interest is the secrecy rate $R_{\mathrm{EJ}}$ under various parameter configurations. For comparison, we also include $R_{\mathrm{GN}}$, $R_{\mathrm{EJ,rand}}$, and $R_{\mathrm{No}}$. Specifically, $R_{\mathrm{EJ,rand}}$ corresponds to the case where the \gls{ris} phase shifts are selected uniformly at random from $[0, 2\pi)$, and only the transmit and jamming precoding matrices are optimized, while $R_{\mathrm{No}}$ denotes the secrecy rate of the no-jammer case, i.e., the classical Multiple-Input Multiple-Output Multiple-Eavesdropper (MIMOME) channel.
	
	In the simulation setup, the large-scale path loss is modeled as 
	\(
	PL(d) = C_0 \left( \frac{d}{d_0} \right)^{-\alpha}
	\),
	where \( C_0 = 10^{-3} \) is the path loss at the reference distance \( d_0 = 1\,\mathrm{m} \), and \( d \) denotes the actual propagation distance between nodes, computed from their predefined position coordinates. The path loss exponent \( \alpha \) is set to 2.2 for links from the BS and jammer to the RIS, 2.5 for links from the RIS to the UE and Eve, and 3.5 for direct links from the BS or jammer to the UE or Eve (bypassing the RIS). {The coordinates (in meters) of the BS, jammer, RIS, UE, and Eve are set to $(0,10)$, $(0,5)$, $(50,5)$, $(49,0)$, and $(60,0)$.}
	
	Owing to the RIS's ability to manipulate reflection phase shifts and concentrate signals toward intended receivers---thus establishing a dominant propagation path---the channels involving the RIS (e.g., BS$\to$RIS$\to$UE, BS$\to$RIS$\to$Eve, jammer$\to$RIS$\to$UE, jammer$\to$RIS$\to$Eve) are modeled as Rician fading channels, $\mathbf{H} = \sqrt{\frac{\beta}{1+\beta}}\mathbf{H}^{\mathrm{LoS}}+\sqrt{\frac{1}{1+\beta}}\mathbf{H}^{\mathrm{NLoS}}$,
	where $\beta$ denotes the Rician factor, while $\mathbf{H}^{\mathrm{LoS}}$ and $\mathbf{H}^{\mathrm{NLoS}}$ represent the deterministic line-of-sight (LoS) component and the stochastic Rayleigh fading/non-LoS (NLoS) component, respectively. {Unless otherwise specified, the Rician factor is set to $\beta=10$ for all RIS-related links.} Assuming that all nodes are equipped with uniform linear arrays (ULAs) with half-wavelength antenna spacing, the LoS component can be expressed as $\mathbf{H}^{\mathrm{LoS}}=\mathbf{a}_{r}\mathbf{a}_{t}^{\mathrm{H}}$, where the transmit and receive steering vectors are given by  
\(\mathbf{a}_{t} = \left[1, e^{j\pi\sin\varphi_{t}},\cdots, e^{j\pi(N_{t}-1)\sin\varphi_{t}}\right]^{\mathrm{T}}\), 
			\(\mathbf{a}_{r} = \left[1, e^{j\pi\sin\varphi_{r}}, \cdots, e^{j\pi(N_{r}-1)\sin\varphi_{r}}\right]^{\mathrm{T}}\),
	with $N_{t}$ and $N_{r}$ denoting the number of transmit and receive antennas (or RIS elements), respectively. The angles $\varphi_{t}$ and $\varphi_{r}$ correspond to the directions of departure and arrival, where
	\(
	\varphi_{t}=\operatorname{atan2}\!\left(y_{r}-y_{t},\,x_{r}-x_{t}\right)\), \(\varphi_{r}=\pi-\varphi_{t}.
	\)
	
	In contrast, channels that bypass the RIS (e.g., BS$\to$UE, BS$\to$Eve, jammer$\to$UE, jammer$\to$Eve) are assumed to follow Rayleigh fading, i.e., the entries of the small-scale fading channel matrices are modeled as i.i.d. random variables drawn from $\mathcal{CN}(0, 1)$.  
	If not otherwise specified, the transmit power constraints are set to $P_1 = P_2 = P$, and the noise power is $\sigma^2 = -100\,\mathrm{dBm}$.\footnote{Normalized noise is adopted in the theoretical analysis for tractability, whereas practical noise levels are used in the simulations to better reflect real-world performance.}
Unless otherwise specified, each curve is averaged over $1000$ random channel realizations.

	\subsection{\texorpdfstring{{MISO Case}}{MISO Case}}
	
\begin{figure}[t]
		\centering
		\includegraphics[width=0.90\columnwidth]{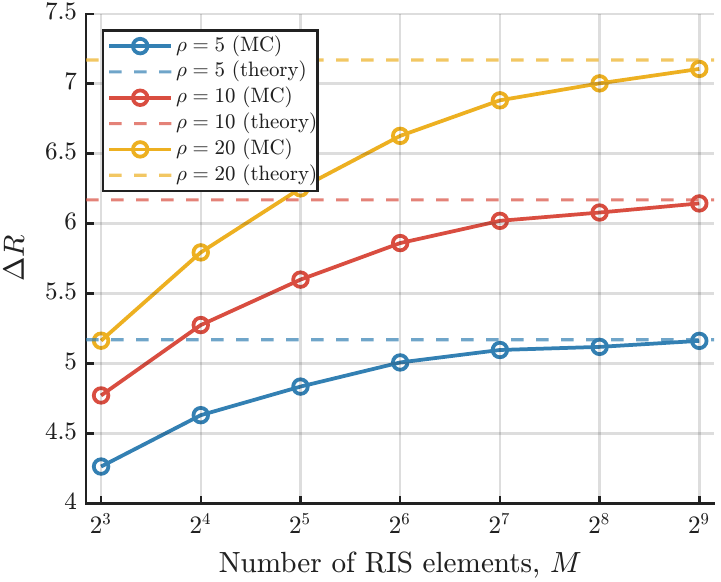}
		\caption{{MISO case: empirical and asymptotic secrecy-rate gaps versus the number of \gls{ris} elements with $(N_{\mathrm{b}},N_{\mathrm{c}},N_{\mathrm{u}},N_{\mathrm{e}})=(4,4,1,1)$, $P_1=30$~dBm, and $\rho=P_2/P_1\in\{5,10,20\}$.}}
		\label{fig:asymptotic_gap}
	\end{figure}

	Fig.~\ref{fig:asymptotic_gap} evaluates the MISO benchmark in Theorem~\ref{thm:main_miso} under the assumptions used in its derivation. The RIS-related links follow mutually independent Rayleigh fading, which corresponds to setting $\beta=0$ in the Rician model above. The direct links are omitted, and the unit-norm precoders are fixed independently of the instantaneous CSI. Using $N_{\rm MC}=5000$ channel realizations, the empirical gap is computed as
	\[
		\begin{aligned}
		\Delta R_M^{\rm MC}
		&=
		\min\left\{
		\frac{1}{N_{\rm MC}}\sum_{n=1}^{N_{\rm MC}}\widetilde R_n,
		\frac{1}{N_{\rm MC}}\sum_{n=1}^{N_{\rm MC}}\widehat R_n
		\right\}\\
		&\quad -
		\frac{1}{N_{\rm MC}}\sum_{n=1}^{N_{\rm MC}}R_{{\rm GN},n}.
		\end{aligned}
	\]
	The empirical curves approach the corresponding asymptotic gaps $\log_2(\rho/\rho_{\rm th})$. At $M=512$, the empirical gaps are $5.162$, $6.144$, and $7.106$, compared with the theoretical values $5.169$, $6.169$, and $7.169$, respectively. This result verifies the phase-aligned benchmark in Theorem~\ref{thm:main_miso}, rather than the output of the joint WMMSE algorithm.
	
\begin{figure}[t]
		\centering
		\includegraphics[width=0.90\columnwidth]{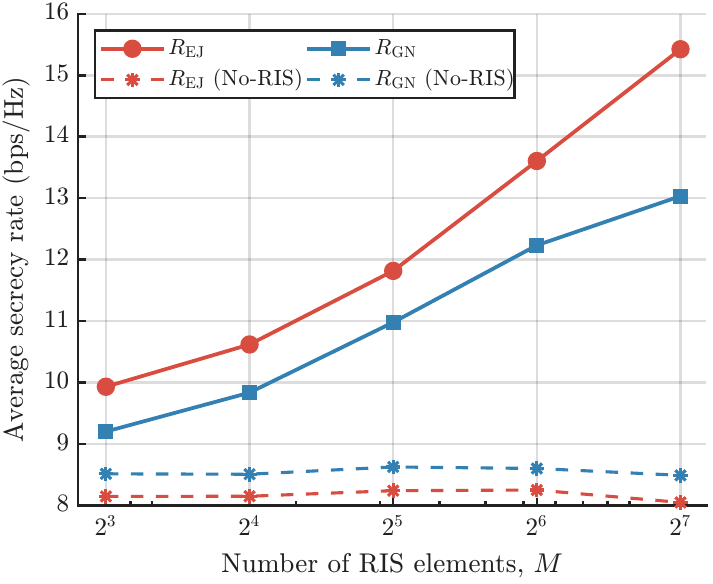}
		\caption{{MISO case: average secrecy rate versus $M$ with $(N_{\mathrm{b}},N_{\mathrm{c}},N_{\mathrm{u}},N_{\mathrm{e}})=(4,4,1,1)$, $P_1=10$~dBm, and $\rho=10$.}}
		\label{fig:miso_wmmse_m}
	\end{figure}

	Fig.~\ref{fig:miso_wmmse_m} evaluates the MISO WMMSE algorithms under the Rician channel model. Without \gls{ris} assistance, \gls{ej} achieves a lower secrecy rate than \gls{gn}, which is consistent with the spatial bottleneck in Theorem~\ref{thm:mi_bounds}. With optimized \gls{ris} phase shifts, both rates increase with $M$, and \gls{ej} outperforms \gls{gn} by more than $15\%$ at $M=128$.

	\subsection{\texorpdfstring{{MIMO Case}}{MIMO Case}}
	
\begin{figure}[t]
		\centering
		\includegraphics[width=0.90\columnwidth]{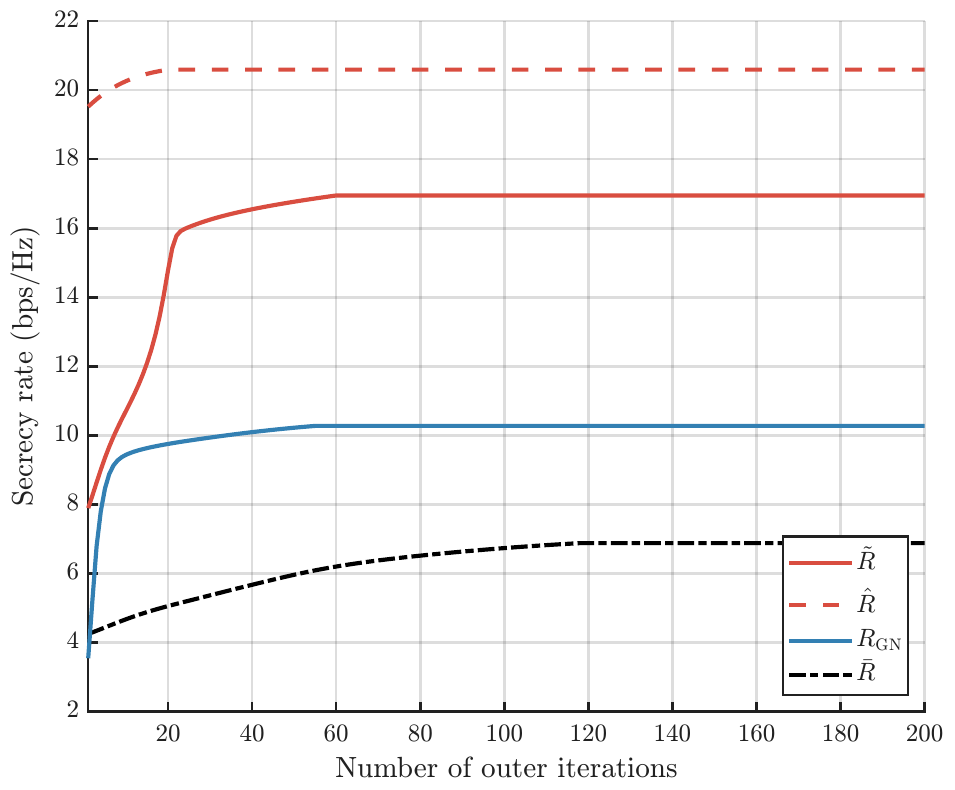}
		\caption{MIMO case: convergence behavior under a single random channel realization with $(N_{\mathrm{b}}, N_{\mathrm{c}}, N_{\mathrm{u}}, N_{\mathrm{e}}) = (2, 4, 4, 4)$, $M=50$, and $P=20$~dBm.}
		\label{fig:convergence}
	\end{figure}

	We next consider the general MIMO case, beginning with the convergence and computational efficiency of the proposed algorithm.
	Fig.~\ref{fig:convergence} shows the secrecy-rate trajectories for the three \gls{ej} candidate subproblems and the \gls{gn} problem under one random channel realization. All four rates increase monotonically and approach stable values. The $\tilde{R}$, $\hat{R}$, and $R_{\mathrm{GN}}$ trajectories reach more than $80\%$ of their respective converged secrecy-rate values within the first 20 outer iterations.
	
\begin{figure}[t]
		\centering
		\includegraphics[width=0.95\columnwidth]{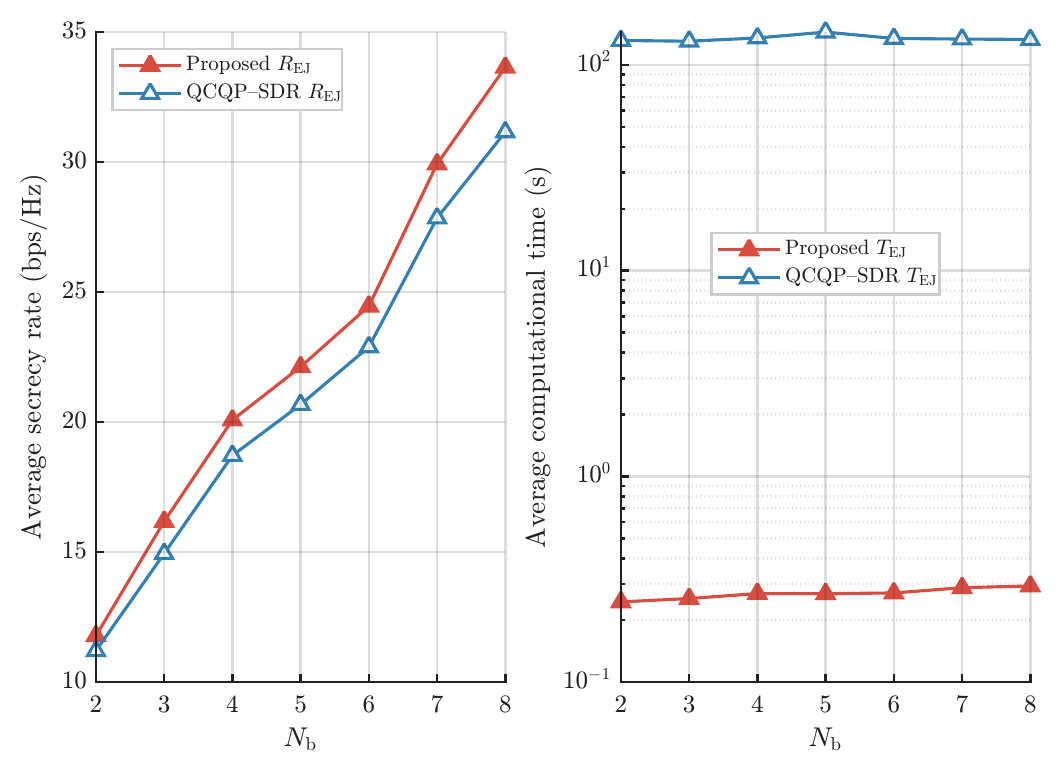}
		\caption{{MIMO case: average secrecy rate and computation time of the proposed EJ-WMMSE and QCQP--SDR algorithms versus $N_{\mathrm b}$ with $N_{\mathrm u}=N_{\mathrm e}=N_{\mathrm c}=4$, $M=50$, and $P=20$~dBm.}}
		\label{fig:ej_wmmse_qcqp_sdr}
	\end{figure}

	In Fig.~\ref{fig:ej_wmmse_qcqp_sdr}, we compare the proposed EJ-WMMSE algorithm with a benchmark based on quadratically constrained quadratic programming (QCQP) and semidefinite relaxation (SDR), referred to as QCQP--SDR. Specifically, the proposed algorithm updates the transmit precoders using closed-form solutions to the precoder QCQPs associated with the candidate subproblems and updates the \gls{ris} phases using MM, whereas the QCQP--SDR benchmark solves the precoder subproblems using CVX and handles the \gls{ris} phase subproblem using SDR. The results, averaged over $200$ random channel realizations, show that the proposed EJ-WMMSE algorithm achieves a higher average secrecy rate for all considered values of $N_{\mathrm b}$, while reducing the computation time by more than two orders of magnitude under the evaluated settings.
	
\begin{figure}[t]
		\centering
		\includegraphics[width=0.90\columnwidth]{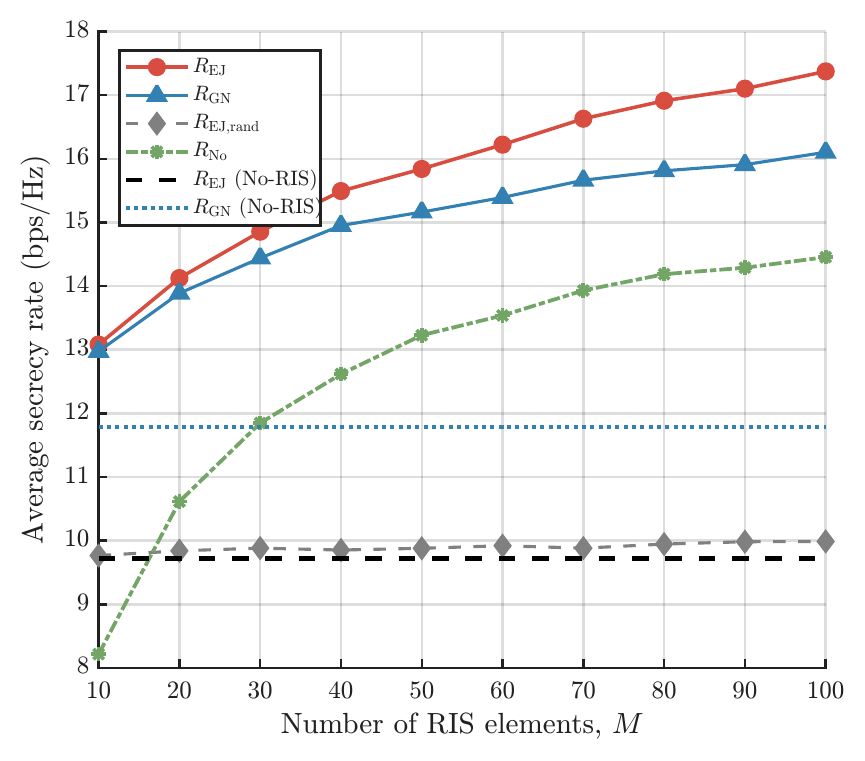}
		\caption{{MIMO case: average secrecy rate versus the number of \gls{ris} elements with $(N_{\mathrm{b}},N_{\mathrm{c}},N_{\mathrm{u}},N_{\mathrm{e}})=(4,4,2,4)$ and $P=20$~dBm.}}
		\label{fig:mimo_rate_vs_m}
	\end{figure}

	Fig.~\ref{fig:mimo_rate_vs_m} evaluates the average secrecy rate versus the number of \gls{ris} elements $M$. Without \gls{ris} assistance, the \gls{ej} scheme achieves a lower secrecy rate than the \gls{gn} scheme. This observation is consistent with the spatial-degradedness bottleneck characterized in Theorem~\ref{thm:mi_bounds}. In the considered antenna configuration, the jammer--Eve channel can span more spatial dimensions than the jammer--UE channel, making the system more susceptible to this bottleneck when Eve's effective jamming-channel gain is sufficiently strong. Nevertheless, the antenna dimensions alone do not guarantee the matrix ordering in Theorem~\ref{thm:mi_bounds}. As $M$ increases, the optimized \gls{ris} reshapes the effective information and jamming channels and alleviates the unfavorable channel relationship. Consequently, both schemes achieve higher secrecy rates, while the \gls{ej} scheme attains a growing advantage over the \gls{gn} scheme. In contrast, $R_{\mathrm{EJ,rand}}$ varies only slightly with $M$ and remains considerably below the optimized result. These results show that the performance improvement mainly arises from the joint optimization of the precoders and \gls{ris} phase shifts, rather than from merely increasing the number of reflecting elements.
	
\begin{figure}[t]
		\centering
		\includegraphics[width=0.90\columnwidth]{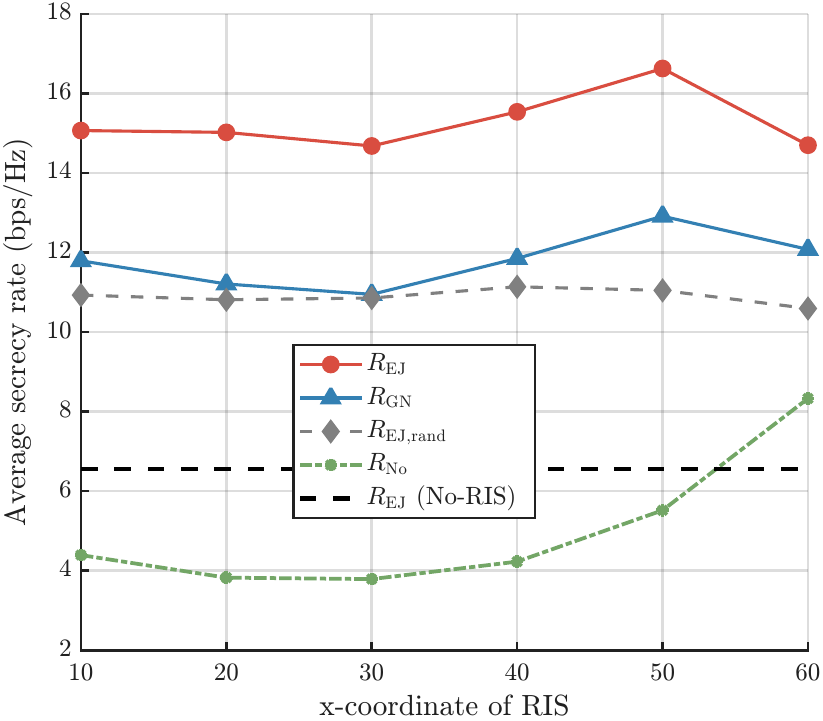}
		\caption{MIMO case: average secrecy rate versus the horizontal \gls{ris} coordinate with $(N_{\mathrm{b}}, N_{\mathrm{c}}, N_{\mathrm{u}}, N_{\mathrm{e}}) = (2, 4, 4, 4)$, $M=50$, and $P=20$~dBm.}
		\label{fig:ris_position}
	\end{figure}

	Furthermore, Fig.~\ref{fig:ris_position} evaluates the \gls{ris} deployment geometry. First, the proposed \gls{ej} scheme shows spatial robustness. For example, at the bottleneck region $x=30$~m, we observe $(R_{\mathrm{EJ}} - R_{\mathrm{No}}) / R_{\mathrm{No}} > 200\%$. Second, the optimal deployment topology shifts. The no-jammer baseline ($R_{\mathrm{No}}$) must deploy the \gls{ris} directly above \gls{eve} ($x=60$~m) to create a spatial null. In contrast, by offloading \gls{eve} suppression to the jammer, the \gls{ej} scheme shifts its maximum secrecy rate point to the vicinity of the \gls{ue} ($x=50$~m). Third, the joint optimization of all spatial variables provides clear array gains over random phase shift configurations. For instance, at $x=50$~m, we observe $(R_{\mathrm{EJ}} - R_{\mathrm{EJ,rand}}) / R_{\mathrm{EJ,rand}} >50\%$. Finally, the \gls{ej} scheme achieves a higher secrecy rate than the \gls{gn} scheme in the evaluated configuration. At the optimal point $x=50$~m, we observe $(R_{\mathrm{EJ}} - R_{\mathrm{GN}}) / R_{\mathrm{GN}} > 25\%$. These results support the usefulness of the optimization framework and the observed advantage of the \gls{ej} scheme in this setting.
	
	{We next examine the antenna-dimension effects by comparing the RIS-free and RIS-assisted MIMO systems.}
	
\begin{figure}[t]
		\centering
		\includegraphics[width=0.90\columnwidth]{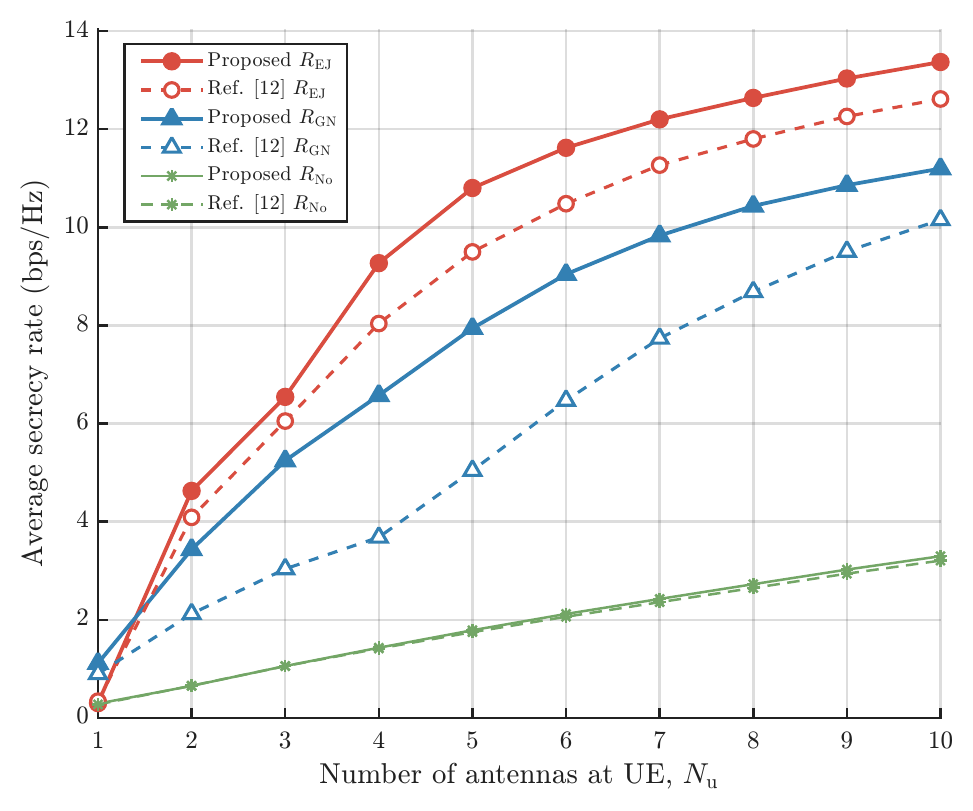}
		\caption{MIMO case without \gls{ris}: average secrecy rate versus $N_{\mathrm{u}}$ with $(N_{\mathrm{b}}, N_{\mathrm{c}}, N_{\mathrm{e}}) = (2, 4, 4)$.}
		\label{fig:noris_vs_nu}
	\end{figure}

\begin{figure}[t]
		\centering
		\includegraphics[width=0.90\columnwidth]{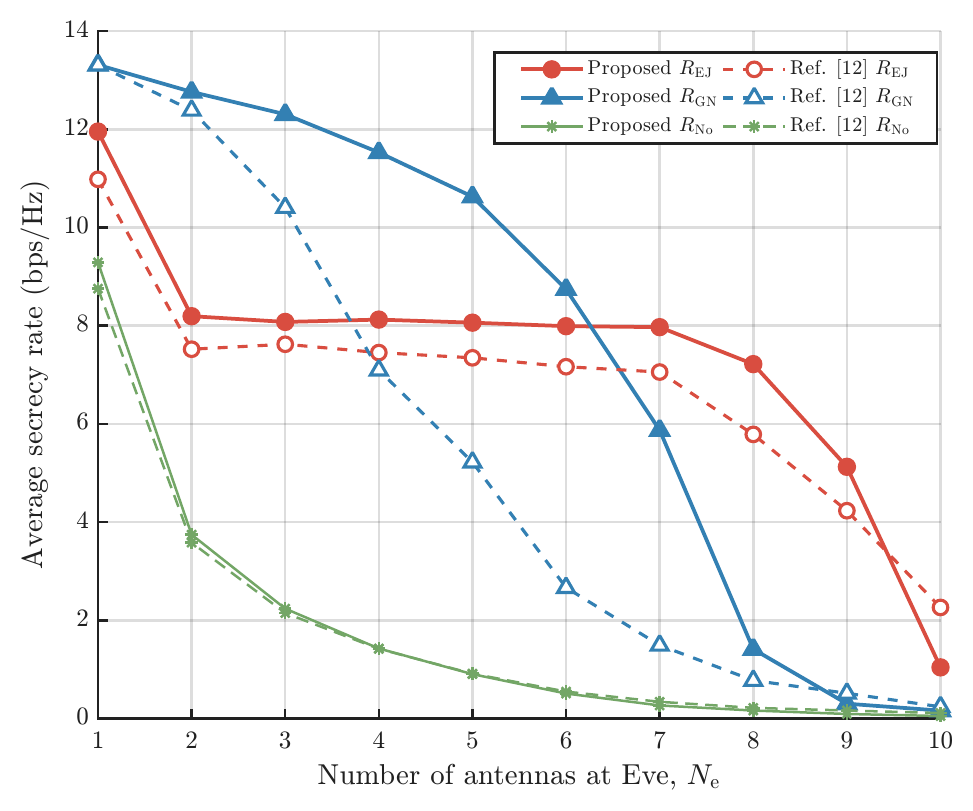}
		\caption{MIMO case without \gls{ris}: average secrecy rate versus $N_{\mathrm{e}}$ with $(N_{\mathrm{b}}, N_{\mathrm{c}}, N_{\mathrm{u}}) = (2, 8, 4)$.}
		\label{fig:noris_vs_ne}
	\end{figure}

	In Fig.~\ref{fig:noris_vs_nu} and Fig.~\ref{fig:noris_vs_ne}, we consider the MIMO case without RIS, where the entries of the channel matrices are i.i.d. $\mathcal{CN}(0,1)$ random variables and $P = 20\,\text{dB}$ under normalized noise. First, a comparison with the SD algorithm in~\cite{xu2024} shows that the proposed algorithm achieves higher secrecy rates under the evaluated antenna configurations.
	Second, a spatial bottleneck exists under the evaluated spatial fading conditions. For instance, in Fig.~\ref{fig:noris_vs_ne}, when $N_{\mathrm{e}}=4$, we observe $(R_{\mathrm{GN}} - R_{\mathrm{EJ}}) / R_{\mathrm{EJ}} > 40\%$. When $N_{\mathrm{e}}$ is small, the jammer's abundant spatial DoFs allow the \gls{gn} scheme to achieve effective zero-forcing at the \gls{ue}. However, the \gls{ej} scheme must maintain sufficient interference power at the \gls{ue} to satisfy the joint-decoding constraint. This decoding constraint results in a performance penalty in such interference-favorable regimes.
	
\begin{figure}[t]
		\centering
		\includegraphics[width=0.90\columnwidth]{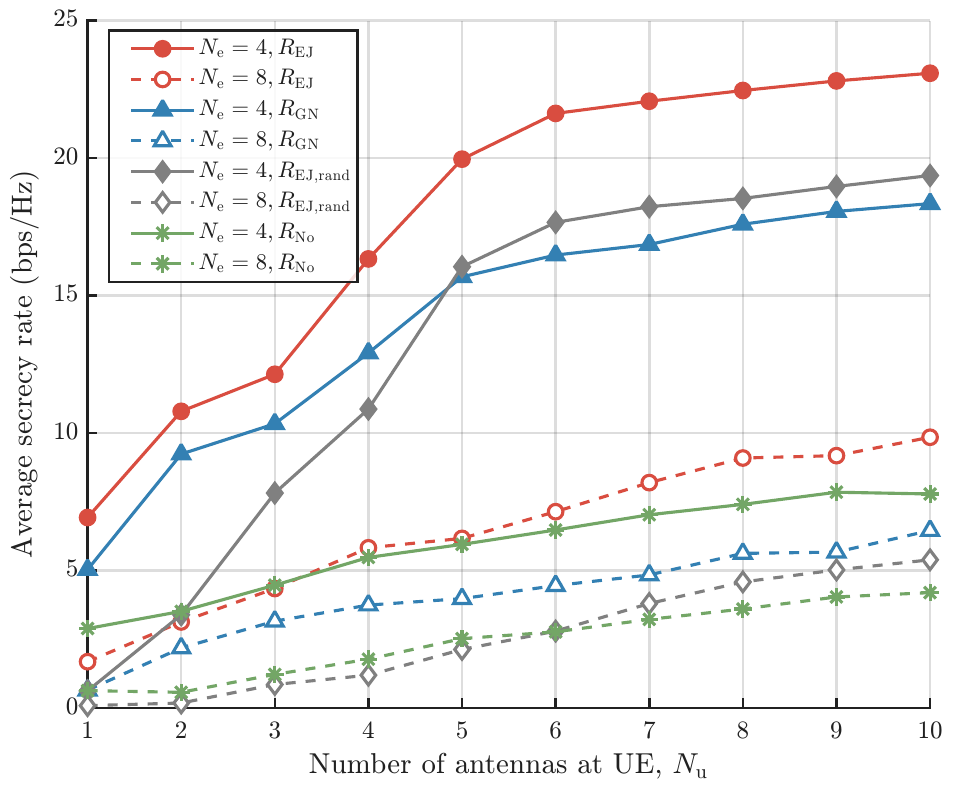}
		\caption{MIMO case: average secrecy rate versus $N_{\mathrm{u}}$ with $N_{\mathrm{e}} \in \{4, 8\}$, $(N_{\mathrm{b}}, N_{\mathrm{c}}) = (2, 4)$, $M=50$, and $P=20$~dBm.}
		\label{fig:ris_vs_nu}
	\end{figure}

\begin{figure}[t]
		\centering
		\includegraphics[width=0.90\columnwidth]{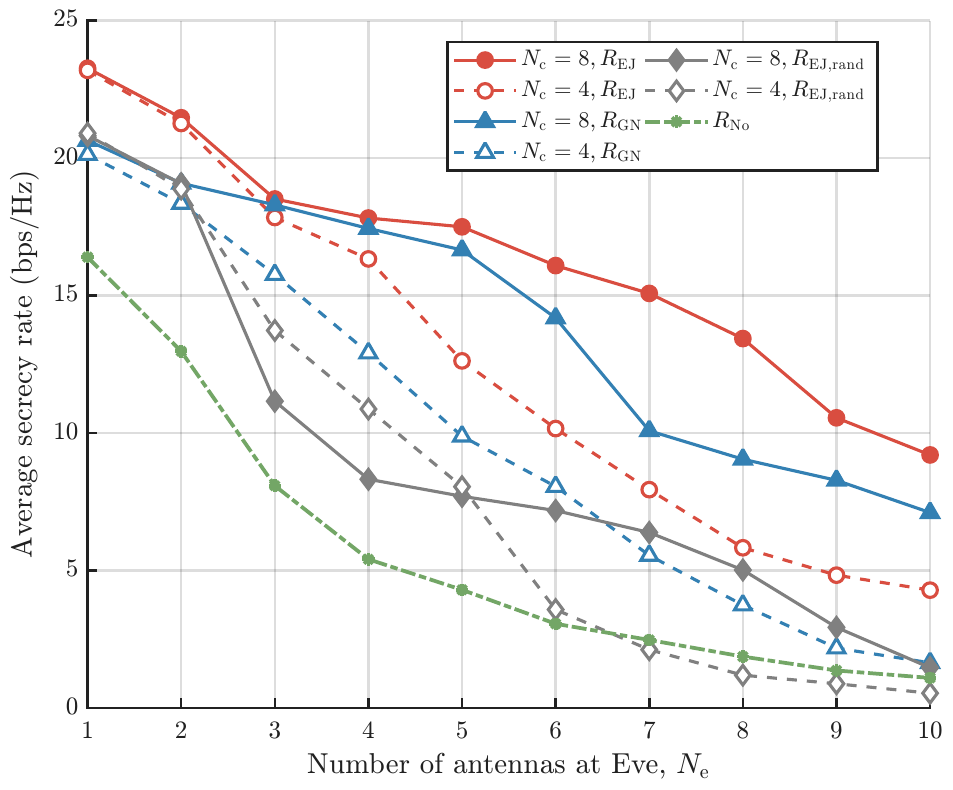}
		\caption{MIMO case: average secrecy rate versus $N_{\mathrm{e}}$ with $N_{\mathrm{c}} \in \{4, 8\}$, $(N_{\mathrm{b}}, N_{\mathrm{u}}) = (2, 4)$, $M=50$, and $P=20$~dBm.}
		\label{fig:ris_vs_ne}
	\end{figure}

	Fig.~\ref{fig:ris_vs_nu} and Fig.~\ref{fig:ris_vs_ne} show the effect of \gls{ris}-based spatial control. First, optimizing the \gls{ris} phase shift matrix is necessary to achieve the array gain. For example, in Fig.~\ref{fig:ris_vs_ne}, when $N_{\mathrm{e}}=4, N_{\mathrm{c}}=8$, we observe $(R_{\mathrm{EJ}} - R_{\mathrm{EJ,rand}}) / R_{\mathrm{EJ,rand}} > 100\%$. Second, in the evaluated configuration, the \gls{ris} helps the \gls{ej} scheme alleviate the aforementioned decoding penalty. As analyzed previously, without the \gls{ris} (Fig.~\ref{fig:noris_vs_ne}), the \gls{ej} scheme is inferior to the \gls{gn} scheme when $N_{\mathrm{e}} \le 6$. However, with \gls{ris} assistance (Fig.~\ref{fig:ris_vs_ne}), the \gls{ej} scheme mitigates this deficit under the same $N_{\mathrm{c}}=8$ configuration. This suggests that the joint optimization framework uses the \gls{ris} to create a favorable equivalent eigenspace, satisfying the joint-decoding constraint at the \gls{ue} while preserving jamming directions toward \gls{eve}.
	Third, the \gls{ej} scheme is robust against DoF depletion. The performance gap varies non-linearly with $N_{\mathrm{e}}$ (Fig.~\ref{fig:ris_vs_ne}). For $N_{\mathrm{c}}=8$, the relative advantage of the \gls{ej} scheme over the \gls{gn} scheme narrows in the lightly loaded regime ($N_{\mathrm{e}} \le 4$), but becomes larger at $N_{\mathrm{e}}=7$, yielding $(R_{\mathrm{EJ}} - R_{\mathrm{GN}}) / R_{\mathrm{GN}} > 45\%$. As $N_{\mathrm{e}}$ approaches $N_{\mathrm{c}}$, the \gls{gn} scheme exhausts its available DoFs to form spatial nulls toward a high-dimensional \gls{eve}, causing its secrecy rate to drop. In contrast, by bypassing the exact nulling requirement, the \gls{ej} scheme avoids DoF depletion, maintaining array gains in the mid-to-high interference regime.
	
	As an additional sensitivity check, we vary the path-loss exponent $\alpha$ for the \gls{ris}--\gls{ue} and \gls{ris}--\gls{eve} links from $2.0$ to $4.0$, while keeping the other parameters unchanged. As the propagation loss becomes more severe, the secrecy rates generally decrease. Nevertheless, under the evaluated configuration, the optimized \gls{ej} scheme remains above the optimized \gls{gn} scheme and the no-jammer baseline throughout the considered range. This numerical observation is specific to the simulated geometry and does not imply a universal ordering between the \gls{ej} scheme and the \gls{gn} scheme.

	\begin{figure}[t]
		\centering
		\includegraphics[width=0.92\columnwidth]{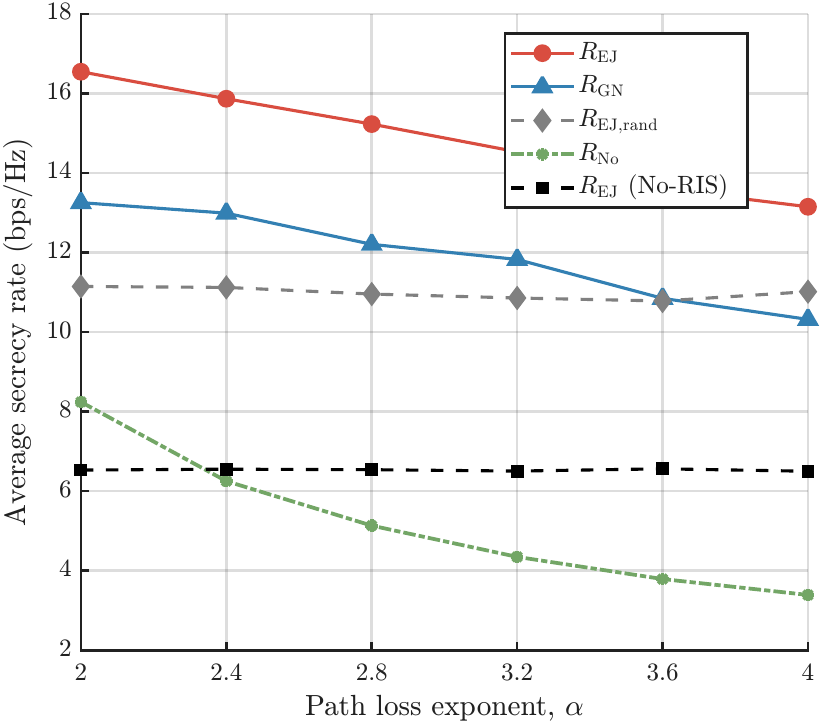}
		\caption{MIMO case: average secrecy rate versus the path-loss exponent $\alpha$ with $(N_{\mathrm{b}}, N_{\mathrm{c}}, N_{\mathrm{u}}, N_{\mathrm{e}}) = (2, 4, 4, 4)$, $M=50$, and $P=20$~dBm.}
		\label{fig:alpha_impact}
	\end{figure}

		\section{Conclusion}\label{conclusion}
		This paper investigated the secure transmission design for an \gls{ris}-assisted MIMO communication system using a cooperative jammer. Starting from the secrecy-rate expressions of the \gls{ej} and \gls{gn} schemes, we identified a spatial channel bottleneck where the \gls{ej} scheme is limited by the joint-decoding constraint under unfavorable jamming-channel conditions. To alleviate this limitation, we introduced an \gls{ris} to reshape the spatial channel topology and formulated a non-convex max-min secrecy rate optimization problem.
		We then developed a WMMSE-based algorithm for joint precoder and RIS phase optimization. We further provided a large-\gls{ris} benchmark showing a positive gap for the jamming-aligned \gls{ej} construction over the signal-aligned \gls{gn} benchmark under the stated conditions.
		Numerical results support the analysis and the proposed design. In the evaluated RIS-assisted settings, the optimized \gls{ej} scheme can outperform the \gls{gn} scheme by transforming part of the jamming leakage into a resolvable signal. These results demonstrate that RIS-assisted channel reconfiguration can alleviate the spatial bottleneck of the \gls{ej} scheme across different antenna configurations.

\appendices
		\section{Proof of Theorem~\ref{thm:mi_bounds}}\label{app:mi_bounds}
		We prove the three properties sequentially based on the chain rule of mutual information and matrix inequalities.
		\textit{Proof of property 1:} Note that the \gls{eve}'s rate term $I(\mathbf{x}_1;\mathbf{y}_{\mathrm{e}})$ cancels out in the difference between $\hat{R}$ and $R_{\mathrm{GN}}$, which gives
		\begin{equation}\label{eq:hat_minus_gn_MI}
			\hat{R} - R_{\mathrm{GN}} = I(\mathbf{x}_1;\mathbf{y}_{\mathrm{u}}\mid \mathbf{x}_2)-I(\mathbf{x}_1;\mathbf{y}_{\mathrm{u}}).
		\end{equation}
		Applying the chain rule of mutual information:
		\begin{equation}
			I(\mathbf{x}_1;\mathbf{y}_{\mathrm{u}}, \mathbf{x}_2)=I(\mathbf{x}_1;\mathbf{x}_2)+I(\mathbf{x}_1;\mathbf{y}_{\mathrm{u}}\mid \mathbf{x}_2).
		\end{equation}
		Since $\mathbf{x}_1$ and $\mathbf{x}_2$ are independent, $I(\mathbf{x}_1;\mathbf{x}_2)=0$, which yields
		\begin{equation}\label{eq:cond_ge_uncond}
			I(\mathbf{x}_1;\mathbf{y}_{\mathrm{u}}\mid \mathbf{x}_2)=I(\mathbf{x}_1;\mathbf{y}_{\mathrm{u}}, \mathbf{x}_2)\ge I(\mathbf{x}_1;\mathbf{y}_{\mathrm{u}}).
		\end{equation}
		
		Substituting \eqref{eq:cond_ge_uncond} into \eqref{eq:hat_minus_gn_MI}, we obtain $\hat{R} \ge R_{\mathrm{GN}}$. Since the non-negative truncation operator $[\cdot]^+$ is monotonically non-decreasing, this inequality is preserved after truncation, leading to the following universal result:
		\begin{equation}
			\hat{R} \ge R_{\mathrm{GN}},\quad \forall~\mathbf{Q}_1\succeq 0,\mathbf{Q}_2\succeq 0.
		\end{equation}
		
		
		\textit{Proof of property 2:} By applying the chain rule of mutual information, we have
		\begin{align}
			\tilde{R} - R_{\mathrm{GN}} 
			&=\Big(I(\mathbf{x}_1,\mathbf{x}_2;\mathbf{y}_{\mathrm{u}})-I(\mathbf{x}_1;\mathbf{y}_{\mathrm{u}})\Big) \nonumber \\
			&\quad -\Big(I(\mathbf{x}_1,\mathbf{x}_2;\mathbf{y}_{\mathrm{e}})-I(\mathbf{x}_1;\mathbf{y}_{\mathrm{e}})\Big)\nonumber \\
			&= I(\mathbf{x}_2;\mathbf{y}_{\mathrm{u}}\mid \mathbf{x}_1)-I(\mathbf{x}_2;\mathbf{y}_{\mathrm{e}}\mid \mathbf{x}_1).
		\end{align}
		For the standard Gaussian linear model, assuming Gaussian inputs and unit-variance noise, the rate difference can be expressed in the following closed form:
		\begin{align}\label{eq:Delta_jam_logdet}
			\tilde{R} - R_{\mathrm{GN}} 
			&= \log\big|\mathbf{I}_{N_{\mathrm{u}}}+\mathbf{H}_2\mathbf{Q}_2\mathbf{H}_2^{\mathrm{H}}\big|
			-\log\big|\mathbf{I}_{N_{\mathrm{e}}}+\mathbf{G}_2\mathbf{Q}_2\mathbf{G}_2^{\mathrm{H}}\big| \nonumber \\
			\stackrel{(a)}{=} \log&\big|\mathbf{I}_{N_{\mathrm{c}}}+\mathbf{F}_2^{\mathrm{H}}\mathbf{H}_2^{\mathrm{H}}\mathbf{H}_2\mathbf{F}_2\big|
			-\log\big|\mathbf{I}_{N_{\mathrm{c}}}+\mathbf{F}_2^{\mathrm{H}}\mathbf{G}_2^{\mathrm{H}}\mathbf{G}_2\mathbf{F}_2\big| \nonumber \\
			&\triangleq \Delta_{\mathrm{jam}}(\mathbf{F}_2, \mathbf{H}_2, \mathbf{G}_2),
		\end{align}
		where $(a)$ follows from substituting $\mathbf{Q}_2 = \mathbf{F}_2 \mathbf{F}_2^{\mathrm{H}}$ and applying the Sylvester determinant identity $|\mathbf{I}+\mathbf{AB}| = |\mathbf{I}+\mathbf{BA}|$. We define $\Delta_{\mathrm{jam}}$ as the \emph{\gls{jdg}}, quantifying how much more ``decodable'' the jamming signal is at the \gls{ue} than at \gls{eve}.
		
		\textit{Proof of property 3:} We first establish the restrictive relationship between $\hat{R}$ and $\tilde{R}$ under the specified spatial degradedness condition. By manipulating the mutual information expressions and applying the Sylvester determinant identity with $\mathbf{Q}_2 = \mathbf{F}_2 \mathbf{F}_2^{\mathrm{H}}$, the difference between the two rates can be formulated as
		\begin{align}
			\hat{R} -& \tilde{R} = \log\big|\mathbf{I}_{N_{\mathrm{c}}} + \mathbf{F}_2^{\mathrm{H}}\mathbf{G}_2^{\mathrm{H}}\mathbf{G}_2\mathbf{F}_2\big|\nonumber \\ - &\log\big|\mathbf{I}_{N_{\mathrm{c}}} + \mathbf{F}_2^{\mathrm{H}}\mathbf{H}_2^{\mathrm{H}} \left(\mathbf{I}_{N_{\mathrm{u}}} + \mathbf{H}_1\mathbf{Q}_1\mathbf{H}_1^{\mathrm{H}}\right)^{-1} \mathbf{H}_2\mathbf{F}_2\big|.
		\end{align}
		
		Since the legitimate signal covariance matrix satisfies $\mathbf{H}_1\mathbf{Q}_1\mathbf{H}_1^{\mathrm{H}} \succeq \mathbf{0}$, we have $\mathbf{I}_{N_{\mathrm{u}}} + \mathbf{H}_1\mathbf{Q}_1\mathbf{H}_1^{\mathrm{H}} \succeq \mathbf{I}_{N_{\mathrm{u}}}$. Inverting this matrix reverses the positive semi-definite partial order, yielding $\left(\mathbf{I}_{N_{\mathrm{u}}} + \mathbf{H}_1\mathbf{Q}_1\mathbf{H}_1^{\mathrm{H}}\right)^{-1} \preceq \mathbf{I}_{N_{\mathrm{u}}}$. Pre- and post-multiplying by $\mathbf{H}_2^{\mathrm{H}}$ and $\mathbf{H}_2$, respectively, provides
		\begin{equation}
			\mathbf{H}_2^{\mathrm{H}} \left(\mathbf{I}_{N_{\mathrm{u}}} + \mathbf{H}_1\mathbf{Q}_1\mathbf{H}_1^{\mathrm{H}}\right)^{-1} \mathbf{H}_2 \preceq \mathbf{H}_2^{\mathrm{H}}\mathbf{H}_2.
		\end{equation}
		
		Given the spatial degradedness condition $\mathbf{G}_2^{\mathrm{H}}\mathbf{G}_2 \succeq \mathbf{H}_2^{\mathrm{H}}\mathbf{H}_2$, the transitivity of the partial order implies
		\begin{equation}
			\mathbf{G}_2^{\mathrm{H}}\mathbf{G}_2 \succeq \mathbf{H}_2^{\mathrm{H}} \left(\mathbf{I}_{N_{\mathrm{u}}} + \mathbf{H}_1\mathbf{Q}_1\mathbf{H}_1^{\mathrm{H}}\right)^{-1} \mathbf{H}_2.
		\end{equation}
		
		Because the function $\log\det(\cdot)$ is monotone with respect to the Loewner order on the cone of positive definite matrices, pre- and post-multiplying the inequality by $\mathbf{F}_2^{\mathrm{H}}$ and $\mathbf{F}_2$ preserves the order, guaranteeing that $\hat{R} - \tilde{R} \ge 0$.
		
		According to the minimum-decision rule of the \gls{ej} scheme, the non-trivial joint-decoding branch for the fixed channels and precoders is $\min\{\hat{R}, \tilde{R}\}=\tilde{R}$. Applying the condition $\mathbf{G}_2^{\mathrm{H}}\mathbf{G}_2 \succeq \mathbf{H}_2^{\mathrm{H}}\mathbf{H}_2$ to \eqref{eq:Delta_jam_logdet} yields $\Delta_{\mathrm{jam}}\le 0$. By Property 2, $\tilde{R}=R_{\mathrm{GN}}+\Delta_{\mathrm{jam}}\le R_{\mathrm{GN}}$. Therefore,
        \begin{equation}
            \min\{\hat{R},\tilde{R}\}\le R_{\mathrm{GN}}
        \end{equation}
        holds pointwise for every feasible precoder whenever the fixed effective channels satisfy the degradedness condition. This statement concerns the joint-decoding branch under the same configuration. The full \gls{ej} expression also contains the no-jamming branch $\bar{R}$, and changing the \gls{ris} phases changes $\mathbf{H}_2$ and $\mathbf{G}_2$ and may change whether the degradedness condition holds. Consequently, this property does not by itself order the separately optimized \gls{ej} and \gls{gn} objectives.

\section{Proof of Theorem~\ref{thm:simo_inversion}}\label{app:simo_asymptotic}

        \begin{proof}
            In the SIMO configuration, the jammer is equipped with a single antenna. Ignoring the direct links, the cascaded jamming channels are $\mathbf{h}_2=\mathbf{G}_{\mathrm{r,u}}\bm{\Phi}\mathbf{g}_{\mathrm{c,r}}$ and $\mathbf{g}_2=\mathbf{G}_{\mathrm{r,e}}\bm{\Phi}\mathbf{g}_{\mathrm{c,r}}$, where $\mathbf{g}_{\mathrm{c,r}}\in\mathbb{C}^{M}$ is the jammer--\gls{ris} channel vector. The entries of $\mathbf{G}_{\mathrm{r,u}}$, $\mathbf{G}_{\mathrm{r,e}}$, and $\mathbf{g}_{\mathrm{c,r}}$ follow the mutually independent Rayleigh model stated in Section~\ref{sec:asymptotic}.

            Construct $\bm{\Phi}^{\dagger}$ to align the cascaded jamming channel at the first \gls{ue} antenna. Its $m$-th diagonal entry is
            \begin{equation}
                [\bm{\Phi}^{\dagger}]_{m,m}=\exp\!\left\{-j\arg\!\left([\mathbf{G}_{\mathrm{r,u}}]_{1,m}[\mathbf{g}_{\mathrm{c,r}}]_m\right)\right\}.
            \end{equation}
            Let $c_{\mathrm{h}}\triangleq \frac{\pi^2}{16}\beta_{\mathrm{c}}\beta_{\mathrm{u}}>0$. By Lemma~\ref{lem:ue_jamming_gain_miso}, the aligned component satisfies
            \begin{equation}
                \frac{|[\mathbf{h}_2]_1|^2}{M^2}\xrightarrow{\mathrm{a.s.}}c_{\mathrm{h}}.
            \end{equation}
            For each remaining \gls{ue} antenna $k\in\{2,\ldots,N_{\mathrm{u}}\}$, the phase configuration acts non-coherently because the rows of $\mathbf{G}_{\mathrm{r,u}}$ are independent. Lemma~\ref{lem:non_coherent_gains_miso} gives $|[\mathbf{h}_2]_k|^2/M=\mathcal{O}_p(1)$, and hence $|[\mathbf{h}_2]_k|^2/M^2\xrightarrow{p}0$. Since $N_{\mathrm{u}}$ is fixed,
            \begin{equation}
                \frac{\|\mathbf{h}_2\|^2}{M^2}\xrightarrow{p}c_{\mathrm{h}}.
            \end{equation}

            The channel $\mathbf{G}_{\mathrm{r,e}}$ is independent of the channels used to construct $\bm{\Phi}^{\dagger}$, so all $N_{\mathrm{e}}$ cascaded components at \gls{eve} combine non-coherently. For each fixed $l$, $|[\mathbf{g}_2]_l|^2/M=\mathcal{O}_p(1)$ and therefore $|[\mathbf{g}_2]_l|^2/M^2\xrightarrow{p}0$. With fixed $N_{\mathrm{e}}$,
            \begin{equation}
                \frac{\|\mathbf{g}_2\|^2}{M^2}\xrightarrow{p}0.
            \end{equation}
            Slutsky's theorem now gives
            \begin{equation}
                \frac{\|\mathbf{h}_2\|^2-\|\mathbf{g}_2\|^2}{M^2}\xrightarrow{p}c_{\mathrm{h}}>0,
            \end{equation}
            which implies $\Pr[\|\mathbf{h}_2\|^2>\|\mathbf{g}_2\|^2]\to 1$. For any fixed nonzero jamming power in the single-antenna-jammer model, this channel-gain ordering is equivalent to $\Delta_{\mathrm{jam}}(\bm{\Phi}^{\dagger})>0$. This completes the proof.
        \end{proof}

\section{Proof of Lemma~\ref{lem:ue_jamming_gain_miso}}\label{app:lemma_coherent}
		\begin{proof}
			Under the jamming-aligned phase shift configuration $\bm{\Phi}_{\mathrm{jam}}$, the cascaded jamming channel at the \gls{ue} is phase-aligned such that its absolute magnitude becomes a coherent sum of the envelopes of individual cascaded elements, given by
			\begin{equation}\label{eq:h2_abs_sum}
				|h_2| = \sum_{m=1}^M \left| [\mathbf{g}_{\mathrm{r,u}}]_m \right| \cdot \left| [\mathbf{g}_{\mathrm{c,r}}]_m \right| \triangleq \sum_{m=1}^M X_m,
			\end{equation}
			where we define $X_m = \left| [\mathbf{g}_{\mathrm{r,u}}]_m \right| \cdot \left| [\mathbf{g}_{\mathrm{c,r}}]_m \right|$. 
			
			First, we evaluate the statistical properties of the individual elements. Under the independent Rayleigh fading assumptions stated in Section~\ref{sec:asymptotic}, $[\mathbf{g}_{\mathrm{r,u}}]_m \sim \mathcal{CN}(0, \beta_{\mathrm{u}})$. Thus, its envelope $\left| [\mathbf{g}_{\mathrm{r,u}}]_m \right|$ follows a Rayleigh distribution, which yields the expectation $\mathbb{E}\left[ \left| [\mathbf{g}_{\mathrm{r,u}}]_m \right| \right] = \frac{\sqrt{\pi}}{2}\sqrt{\beta_{\mathrm{u}}}$. Since the cascaded channels are independent, the expectation of the product $X_m$ is $\mathbb{E}[X_m] = \frac{\pi}{4}\sqrt{\beta_{\mathrm{u}} \beta_{\mathrm{c}}} = \frac{\pi}{4}\sqrt{\beta_{\mathrm{c}} \beta_{\mathrm{u}}}$. The second moment of $X_m$ is $\mathbb{E}[X_m^2] = \beta_{\mathrm{c}} \beta_{\mathrm{u}}$, yielding a finite variance $\mathrm{Var}(X_m) = \beta_{\mathrm{c}} \beta_{\mathrm{u}} \left(1 - \frac{\pi^2}{16}\right)$.
			
			Next, by rewriting the cascaded channel as its deterministic mean plus a zero-mean fluctuation term, we have $|h_2| = M \frac{\pi}{4}\sqrt{\beta_{\mathrm{c}} \beta_{\mathrm{u}}} + Z_M$, where the fluctuation $Z_M \triangleq \sum_{m=1}^M (X_m - \mathbb{E}[X_m])$ is the sum of i.i.d. random variables with finite variance. 
			
			Normalizing the channel power by $M^2$, we obtain the expansion
			\begin{equation}\label{eq:h2_normalized}
				\frac{|h_2|^2}{M^2} = \frac{\pi^2}{16} \beta_{\mathrm{c}} \beta_{\mathrm{u}} + \frac{\pi}{2}\sqrt{\beta_{\mathrm{c}} \beta_{\mathrm{u}}} \left(\frac{Z_M}{M}\right) + \left(\frac{Z_M}{M}\right)^2.
			\end{equation}
			By the SLLN, the sample mean of the zero-mean fluctuations converges almost surely to zero, i.e., $Z_M/M \xrightarrow{a.s.} 0$ as $M \to \infty$. Consequently, the cross term and the residual term in \eqref{eq:h2_normalized} vanish almost surely. This establishes the channel hardening effect, yielding $\frac{1}{M^2}|h_2|^2 \xrightarrow{a.s.} \frac{\pi^2}{16} \beta_{\mathrm{c}} \beta_{\mathrm{u}}$ and completing the proof.
		\end{proof}
		
		\section{Proof of Lemma~\ref{lem:non_coherent_gains_miso}}\label{app:lemma_noncoherent}
		\begin{proof}
			We prove the scaling law for $g_2$; the remaining channels follow the same logic due to the structural symmetry and mutual independence.
			
			Recall the scalar cascaded jamming channel at \gls{eve}, expressed as
			\[
			g_2 = \mathbf{g}_{\mathrm{r,e}}^{\mathrm{H}} \bm{\Phi}_{\mathrm{jam}} \mathbf{g}_{\mathrm{c,r}} = \sum_{m=1}^M [\mathbf{g}_{\mathrm{r,e}}^*]_m [\bm{\Phi}_{\mathrm{jam}}]_{m,m} [\mathbf{g}_{\mathrm{c,r}}]_m \triangleq \sum_{m=1}^M Y_m.
			\]
			Under the independent Rayleigh fading assumptions stated in Section~\ref{sec:asymptotic}, $\mathbf{g}_{\mathrm{r,e}}$ is independent of both $\mathbf{g}_{\mathrm{r,u}}$ and $\mathbf{g}_{\mathrm{c,r}}$. Because $\bm{\Phi}_{\mathrm{jam}}$ is designed using $\mathbf{g}_{\mathrm{r,u}}$ and $\mathbf{g}_{\mathrm{c,r}}$, the phase shift $[\bm{\Phi}_{\mathrm{jam}}]_{m,m}$ is statistically independent of $[\mathbf{g}_{\mathrm{r,e}}^*]_m$.
			
			Therefore, $Y_m$ is the product of independent random variables. The expectation is evaluated as
			\begin{equation}
				\mathbb{E}[Y_m] = \mathbb{E}\left[ [\mathbf{g}_{\mathrm{r,e}}^*]_m \right] \cdot \mathbb{E}\left[ [\bm{\Phi}_{\mathrm{jam}}]_{m,m} [\mathbf{g}_{\mathrm{c,r}}]_m \right] = 0,
			\end{equation}
			since $\mathbb{E}\left[ [\mathbf{g}_{\mathrm{r,e}}^*]_m \right] = 0$. The variance of $Y_m$ is given by
			\begin{align}
				\mathbb{E}[|Y_m|^2] &= \mathbb{E}\left[ \left| [\mathbf{g}_{\mathrm{r,e}}^*]_m \right|^2 \right] \cdot \mathbb{E}\left[ \left| [\bm{\Phi}_{\mathrm{jam}}]_{m,m} [\mathbf{g}_{\mathrm{c,r}}]_m \right|^2 \right] \nonumber \\
				&= \beta_{\mathrm{e}} \cdot \mathbb{E}\left[ \left| [\mathbf{g}_{\mathrm{c,r}}]_m \right|^2 \right] = \beta_{\mathrm{e}} \beta_{\mathrm{c}} = \beta_{\mathrm{c}} \beta_{\mathrm{e}},
			\end{align}
			where we utilize the unit-modulus property $|[\bm{\Phi}_{\mathrm{jam}}]_{m,m}|^2 = 1$. 
			
			Since $\{Y_m\}_{m=1}^M$ constitutes a sequence of i.i.d. complex random variables with zero mean and variance $\beta_{\mathrm{c}} \beta_{\mathrm{e}}$, we can apply the Lindeberg-Levy CLT. As $M \to \infty$, we have
			\begin{equation}
				\frac{1}{\sqrt{M}} g_2 = \frac{1}{\sqrt{M}} \sum_{m=1}^M Y_m \xrightarrow{d} \mathcal{CN}(0, \beta_{\mathrm{c}} \beta_{\mathrm{e}}).
			\end{equation}
			By the CMT, taking the squared magnitude of both sides preserves the convergence in distribution. The squared envelope of a circularly symmetric complex Gaussian random variable follows an exponential distribution. Thus, we obtain
			\begin{equation}
				\frac{1}{M} |g_2|^2 \xrightarrow{d} \beta_{\mathrm{c}} \beta_{\mathrm{e}} \cdot \mathrm{Exp}(1),
			\end{equation}
			which implies $|g_2|^2 ={\mathcal{O}_p}(M)$.
			
			For $h_1 = \sum_{m=1}^M [\mathbf{h}_{\mathrm{r,u}}^*]_m [\bm{\Phi}_{\mathrm{jam}}]_{m,m} [\mathbf{h}_{\mathrm{b,r}}]_m$, the independence of the zero-mean vector $\mathbf{h}_{\mathrm{b,r}}$ yields zero-mean summands with variance $\beta_{\mathrm{b}} \beta_{\mathrm{u}}$. Following the same CLT-CMT derivation as $g_2$, we have $\frac{1}{M} |h_1|^2 \xrightarrow{d} \beta_{\mathrm{b}} \beta_{\mathrm{u}} V_{h_1}$ with $V_{h_1} \sim \mathrm{Exp}(1)$.
			Similarly, $g_1 = \mathbf{h}_{\mathrm{r,e}}^{\mathrm{H}} \bm{\Phi}_{\mathrm{jam}} \mathbf{h}_{\mathrm{b,r}}$ consists of i.i.d. zero-mean summands with variance $\beta_{\mathrm{b}} \beta_{\mathrm{e}}$ due to the mutual independence of Rayleigh channels $\mathbf{h}_{\mathrm{r,e}}$ and $\mathbf{h}_{\mathrm{b,r}}$. Thus, it follows that $\frac{1}{M} |g_1|^2 \xrightarrow{d} \beta_{\mathrm{b}} \beta_{\mathrm{e}} V_{g_1}$ with $V_{g_1} \sim \mathrm{Exp}(1)$. The vector of normalized sums obeys the multivariate complex central limit theorem. Its limiting cross-covariances and pseudo-covariances vanish because the receive-side channel vectors are mutually independent and circularly symmetric. Hence the limiting complex Gaussian components, and therefore $V_{g_2}$, $V_{h_1}$, and $V_{g_1}$, are mutually independent.
		\end{proof}

	\bibliographystyle{IEEEtran}
	\bibliography{IEEEabrv,refs}

\end{document}

%% file: macros.tex
\long\def\comment#1{}

\newcommand{\Fm}{{\mathbf F}}

\renewcommand{\det}{{\hbox{det}}}

\renewcommand{\arg}{{\hbox{arg}}}

\newtheorem{cor}{Corollary}
\newtheorem{lem}{Lemma}

\providecommand{\definitionname}{Definition}

%% file: IEEEabrv.bib
@STRING{IEEE_J_COML       = "{IEEE} Commun. Lett."}

@STRING{IEEE_J_JSAC       = "{IEEE} J. Sel. Areas Commun."}

@STRING{IEEE_J_COM        = "{IEEE} Trans. Commun."}

@STRING{IEEE_J_WCOM       = "{IEEE} Trans. Wireless Commun."}

@STRING{IEEE_J_WCOML      = "{IEEE} Wireless Commun. Lett."}

@STRING{IEEE_J_IFS        = "{IEEE} Trans. Inf. Forensics Security"}

@STRING{IEEE_J_IT         = "{IEEE} Trans. Inf. Theory"}

@STRING{IEEE_J_IOT        = "{IEEE} Internet Things J."}


%% file: refs.bib
@article{arzykulovArtificialNoiseRISaided2023,
  title = {Artificial Noise and {{RIS-aided}} Physical Layer Security: {{Optimal RIS}} Partitioning and Power Control},
  shorttitle = {Artificial Noise and {{RIS-aided}} Physical Layer Security},
  author = {Arzykulov, Sultangali and Celik, Abdulkadir and Nauryzbayev, Galymzhan and Eltawil, Ahmed M.},
  year = {2023},
  month = jun,
  journal = IEEE_J_WCOML ,
  volume = {12},
  number = {6},
  pages = {992--996},
  issn = {2162-2345},
  doi = {10.1109/LWC.2023.3256001},
  urldate = {2025-05-30}
}

@article{illiEnhancingPhysicalLayer2024,
  title = {Enhancing Physical Layer Security with Reconfigurable Intelligent Surfaces and Friendly Jamming: {{A}} Secrecy Analysis},
  author = {Illi, Elmehdi and Qaraqe, Marwa and El Bouanani, Faissal and {Al-Kuwari}, Saif},
  year = {2024},
  month = may,
  journal = {Computer Communications},
  volume = {221},
  pages = {106--119},
  issn = {01403664},
}

@article{tekin2008general,
  author  = {Tekin, E. and Yener, A.},
  title   = {The General Gaussian Multiple-Access and Two-Way Wiretap Channels: Achievable Rates and Cooperative Jamming},
  journal = {IEEE Trans. Inf. Theory},
  volume  = {54},
  number  = {6},
  pages   = {2735--2751},
  month   = jun,
  year    = {2008}
}

@article{hu2018cooperative,
  author  = {Hu, L. and Wen, H. and Wu, B. and Tang, J. and Pan, F. and Liao, R.-F.},
  title   = {Cooperative-Jamming-Aided Secrecy Enhancement in Wireless Networks with Passive Eavesdroppers},
  journal = {IEEE Trans. Veh. Tech.},
  volume  = {67},
  number  = {3},
  pages   = {2108--2117},
  month   = mar,
  year    = {2018}
}

@article{chu2014secrecy,
  author  = {Chu, Z. and Cumanan, K. and Ding, Z. and Johnston, M. and Le Goff, S.~Y.},
  title   = {Secrecy Rate Optimizations for a {MIMO }Secrecy Channel with a Cooperative Jammer},
  journal = {IEEE Trans. Veh. Tech.},
  volume  = {64},
  number  = {5},
  pages   = {1833--1847},
  month   = may,
  year    = {2015}
}

@article{li2014secrecy,
  author  = {Li, L. and Chen, Z. and Fang, J.},
  title   = {On Secrecy Capacity of {Gaussian} Wiretap Channel Aided by a Cooperative Jammer},
  journal = {IEEE Sig. Process. Lett.},
  volume  = {21},
  number  = {11},
  pages   = {1356--1360},
  month   = nov,
  year    = {2014}
}

@article{chenPhysicalLayerSecurity2024,
  title = {Physical Layer Security Improvement for Hybrid {{RIS-assisted MIMO}} Communications},
  author = {Chen, Zhen and Guo, Yeyong and Zhang, Peichang and Jiang, Hao and Xiao, Yuhang and Huang, Lei},
  year = {2024},
  month = nov,
  journal = IEEE_J_COML   ,
  volume = {28},
  number = {11},
  pages = {2493--2497},
  issn = {1558-2558},
  doi = {10.1109/LCOMM.2024.3427010},
  urldate = {2025-05-30}
}

@article{cuiSecureWirelessCommunication2019,
  title = {Secure Wireless Communication via Intelligent Reflecting Surface},
  author = {Cui, Miao and Zhang, Guangchi and Zhang, Rui},
  year = {2019},
  month = oct,
  journal = {IEEE Wireless Commun. Lett.},
  volume = {8},
  number = {5},
  pages = {1410--1414},
  issn = {2162-2337, 2162-2345},
  doi = {10.1109/LWC.2019.2919685},
  urldate = {2025-04-22},
  copyright = {https://ieeexplore.ieee.org/Xplorehelp/downloads/license-information/IEEE.html},
  langid = {english}
}

@article{guanIntelligentReflectingSurface2020a,
  title = {Intelligent Reflecting Surface Assisted Secrecy Communication: {{Is}} Artificial Noise Helpful or Not?},
  shorttitle = {Intelligent Reflecting Surface Assisted Secrecy Communication},
  author = {Guan, Xinrong and Wu, Qingqing and Zhang, Rui},
  year = {2020},
  month = jun,
  journal = {IEEE Wireless Commun. Lett.},
  volume = {9},
  number = {6},
  pages = {778--782},
  issn = {2162-2337, 2162-2345},
  doi = {10.1109/LWC.2020.2969629},
  urldate = {2025-05-12},
  copyright = {https://ieeexplore.ieee.org/Xplorehelp/downloads/license-information/IEEE.html},
  langid = {english}
}

@article{kaurSurveyReconfigurableIntelligent2024,
  title   = {A Survey on Reconfigurable Intelligent Surface for Physical Layer Security of Next-Generation Wireless Communications},
  author  = {Kaur, Ravneet and Bansal, Bajrang and Majhi, Sudhan and Jain, Sandesh and Huang, Chongwen and Yuen, Chau},
  journal = {IEEE Open J. Veh. Technol.},
  volume  = {5},
  pages   = {172--199},
  month   = Jan,
  year    = 2024,
  doi     = {10.1109/OJVT.2023.3348658},
  issn    = {2644-1330}
}

@inproceedings{wuMIMOSecureCommunication2022,
  title = {{{MIMO}} Secure Communication with Reconfigurable Intelligent Surface: {{Finite-alphabet}} Inputs},
  shorttitle = {{{MIMO}} Secure Communication with Reconfigurable Intelligent Surface},
booktitle = {Proc. Int. Conf. on Wirel. Commun. and Signal Process. (WCSP)},
 address = {Nanjing, China},
  author = {Wu, Yingjie and Wang, Shilian and Luo, Junshan and Chen, Weiyu},
  year = {2022},
  month = nov,
  pages = {950--954},
  doi = {10.1109/WCSP55476.2022.10039314},
  urldate = {2025-05-30},
}

@inproceedings{yuEnablingSecureWireless2019,
  title = {Enabling Secure Wireless Communications via Intelligent Reflecting Surfaces},
 booktitle = {Proc. IEEE Global Commun. Conf. (GLOBECOM)},
  author = {Yu, Xianghao and Xu, Dongfang and Schober, Robert},
  year = {2019},
  month = dec,
  pages = {1--6},
 
  doi = {10.1109/GLOBECOM38437.2019.9014322},
  urldate = {2025-04-22},
  copyright = {https://ieeexplore.ieee.org/Xplorehelp/downloads/license-information/IEEE.html},
  isbn = {978-1-7281-0962-6},
  langid = {english}
}

@article{zhangPhysicalLayerSecurity2021,
  title   = {Physical Layer Security Enhancement with Reconfigurable Intelligent Surface-Aided Networks},
  author  = {Zhang, Jiayi and Du, Hongyang and Sun, Qiang and Ai, Bo and Ng, Derrick Wing Kwan},
  journal = IEEE_J_IFS,
  volume  = {16},
  number  = {7},
  pages   = {3480--3495},
  month   = may,
  year    = 2021,
  doi     = {10.1109/TIFS.2021.3083409},
  issn    = {1556-6021}
}

@inproceedings{chuRobustBeamformingTechniques2015,
  title     = {Robust Beamforming Techniques for {MISO} Secrecy Communication with a Cooperative Jammer},
  author    = {Chu, Zheng and Johnston, Martin and Le Goff, Stephane},
  booktitle = {Proc. IEEE Veh. Technol. Conf. (VTC Spring)},
  month     = may,
  year      = 2015,
  address   = {Glasgow, UK},
  pages     = {1--5},
  doi       = {10.1109/VTCSpring.2015.7146065},
  issn      = {1550-2252}
}

@article{wuSecureMassiveMIMO2016,
  title = {Secure Massive {{MIMO}} Transmission with an Active Eavesdropper},
  author = {Wu, Yongpeng and Schober, Robert and Ng, Derrick Wing Kwan and Xiao, Chengshan and Caire, Giuseppe},
  year = {2016},
  month = jul,
  journal = IEEE_J_IT ,
  volume = {62},
  number = {7},
  pages = {3880--3900},
  issn = {1557-9654},
  doi = {10.1109/TIT.2016.2569118},
  urldate = {2025-07-16},
  
}

@inproceedings{wuIntelligentReflectingSurface2018,
  title = {Intelligent Reflecting Surface Enhanced Wireless Network: {{Joint}} Active and Passive Beamforming Design},
  shorttitle = {Intelligent Reflecting Surface Enhanced Wireless Network},
 booktitle = {Proc. IEEE Global Commun. Conf. (GLOBECOM)},
  author = {Wu, Qingqing and Zhang, Rui},
  year = {2018},
  month = dec,
  pages = {1--6},
  address = {Abu Dhabi, United Arab Emirates},
  doi = {10.1109/GLOCOM.2018.8647620},
  urldate = {2025-04-22},
  isbn = {978-1-5386-4727-1},
  langid = {english}
}

@ARTICLE{RISMIMO1,
  author={Hong, Sheng and Pan, Cunhua and Ren, Hong and Wang, Kezhi and Nallanathan, Arumugam},
  journal=IEEE_J_COM, 
  title={Artificial-Noise-Aided Secure {MIMO} Wireless Communications via Intelligent Reflecting Surface}, 
  year={2020},
  month=dec,
  volume={68},
  number={12},
  pages={7851-7866},
  doi={10.1109/TCOMM.2020.3024621}
}

@ARTICLE{xu2024,
  author={Xu, Hao and Wong, Kai-Kit and Xu, Yinfei and Caire, Giuseppe},
  journal=IEEE_J_COM, 
  title={Coding-Enhanced Cooperative Jamming for Secret Communication: The {MIMO} Case}, 
  year={2024},
  month = may,
  volume={72},
  number={5},
  pages={2746-2761},
  doi={10.1109/TCOMM.2024.3355308}
}

@inproceedings{yassaeeMultipleAccessWiretap2010,
  title = {Multiple Access Wiretap Channels with Strong Secrecy},
  booktitle = {Proc. IEEE Inf. Theory Workshop},
  author = {Yassaee, Mohammad Hossein and Aref, Mohammad Reza},
  year = {2010},
  address={Dublin, Ireland},
  month = aug,
  pages = {1--5},
  doi = {10.1109/CIG.2010.5592953}
}

@article{ref10,
  author = {Liu, Yiliang and Chen, Hsiao-Hwa and Wang, Liangmin},
  title = {Physical Layer Security for next Generation Wireless Networks: Theories, Technologies, and Challenges},
  journal = {IEEE Commun. Surv. Tut.},
  year = {2017},
  month = feb,
  volume = {19},
  number = {1},
  pages = {347--376},
  issn = {1553-877X},
  doi = {10.1109/COMST.2016.2598968}
}

@ARTICLE{direnzoSmartRadioEnvironments2020,
  author = {Di Renzo, Marco and Zappone, Alessio and Debbah, Merouane and Alouini, Mohamed-Slim and Yuen, Chau and de Rosny, Julien and Tretyakov, Sergei},
  title = {Smart Radio Environments Empowered by Reconfigurable Intelligent Surfaces: How It Works, State of Research, and the Road Ahead},
  journal = IEEE_J_JSAC ,
  year = {2020},
  month = nov,
  volume = {38},
  number = {11},
  pages = {2450--2525},
  issn = {1558-0008},
  doi = {10.1109/JSAC.2020.3007211}
}

@ARTICLE{7079465,
  author={Yang, Nan and Yan, Shihao and Yuan, Jinhong and Malaney, Robert and Subramanian, Ramanan and Land, Ingmar},
  journal=IEEE_J_COM, 
  title={Artificial Noise: Transmission Optimization in Multi-Input Single-Output Wiretap Channels}, 
  year={2015},
  volume={63},
  number={5},
  pages={1771-1783},
  month=may, 
  doi={10.1109/TCOMM.2015.2419634}
}

@article{ref41,
  author = {Q. Shi and M. Razaviyayn and Z.-Q. Luo and C. He},
  title = {An iteratively weighted MMSE approach to distributed sum-utility maximization for a {MIMO} interfering broadcast channel},
  journal = {IEEE Trans. Signal Process.},
  volume = {59},
  number = {9},
  pages = {4331--4340},
  year = {2011},
  month=sep,  
}

@INPROCEEDINGS{xu16,
  author = {X. Tang and R. Liu and P. Spasojevic and H. V. Poor},
  booktitle = {Proc. IEEE Int. Symp. Inf. Theory (ISIT)}, 
  title = {The {G}aussian Wiretap Channel with a Helping Interferer},
  year = {2008},
  month = jul,  
  address = {Toronto, ON, Canada},
  pages = {389--393},
  doi = {10.1109/ISIT.2008.4595014}
}

@ARTICLE{xu17,
  author = {X. Tang and R. Liu and P. Spasojević and H. V. Poor},
  journal = {IEEE Trans. Inf. Theory}, 
  title = {Interference Assisted Secret Communication}, 
  year = {2011},
  month = may,
  volume = {57},
  number = {5},
  pages = {3153--3167},
  doi = {10.1109/TIT.2011.2121450}
}

@ARTICLE{xu18,
  author = {X. He and A. Yener},
  journal = {IEEE Trans. Inf. Theory}, 
  title = {Providing Secrecy with Structured Codes: {T}wo-{U}ser {G}aussian Channels}, 
  year = {2014},
  month = apr, 
  volume = {60},
  number = {4},
  pages = {2121--2138},
  doi = {10.1109/TIT.2014.2298132}
}

@ARTICLE{shi2015secure,
  author = {Q. Shi and W. Xu and J. Wu and E. Song and Y. Wang},  
  journal = {IEEE Trans. Wireless Commun.},  
  title = {Secure Beamforming for {MIMO} Broadcasting With Wireless Information and Power Transfer}, 
  year = {2015},
  month = may,  
  volume = {14},
  number = {5},
  pages = {2841--2853},  
  doi = {10.1109/TWC.2015.2395414}
}

@article{wangEnergyEfficientRobust2021,
  title = {Energy Efficient Robust Beamforming and Cooperative Jamming Design for {{IRS-assisted MISO}} Networks},
  author = {Wang, Qun and Zhou, Fuhui and Hu, Rose Qingyang and Qian, Yi},
  year = {2021},
  month = apr,
  journal = IEEE_J_WCOM,
  volume = {20},
  number = {4},
  pages = {2592--2607},
  issn = {1536-1276, 1558-2248},
  doi = {10.1109/TWC.2020.3043325},
  urldate = {2025-04-21},
  copyright = {https://ieeexplore.ieee.org/Xplorehelp/downloads/license-information/IEEE.html},
  langid = {english}
}

@article{zhang2023,
  author  = {Zhang, T. and Wen, H. and Jiang, Y. and Tang, J.},
  title   = {Deep-Reinforcement-Learning-Based IRS for Cooperative Jamming Networks Under Edge Computing},
  journal = IEEE_J_IOT,
  volume  = {10},
  number  = {10},
  pages   = {8996--9006},
  month   = may,
  year    = {2023},
  doi     = {10.1109/JIOT.2022.3232587}
}

@ARTICLE{Wen2025,
  author={Wen, Yingkun and Wang, Fengshuan and Wang, Hui-Ming and Li, Junhuai and Qian, Jin and Wang, Kan and Wang, Huaijun},
  journal=IEEE_J_COM, 
  title={Cooperative Jamming Aided Secure Communication for {RIS} Enabled Symbiotic Radio Systems}, 
  year={2025},
  volume={73},
mouth=may,
  number={5},
  pages={2936-2949},
  doi={10.1109/TCOMM.2024.3481038}}

@book{zhang2017matrix,
  author    = {X. D. Zhang},
  title     = {Matrix Analysis and Applications},
  publisher = {Cambridge University Press},
  year      = {2017}
}

@ARTICLE{pan2019multicell,
  author={Pan, Cunhua and Ren, Hong and Wang, Kezhi and Xu, Wei and Elkashlan, Maged and Nallanathan, Arumugam and Hanzo, Lajos},
  journal=IEEE_J_WCOM, 
  title={Multicell {MIMO} Communications Relying on Intelligent Reflecting Surfaces}, 
  year={2020},
  volume={19},
  number={8},
  pages={5218--5233},
  month=aug,
  doi={10.1109/TWC.2020.2990766}
}

@inproceedings{xu2023achievable,
	title={Achievable Region of the {$K$}-User {MAC} Wiretap Channel Under Strong Secrecy},
	author={Xu, Hao and Wong, Kai-Kit and Caire, Giuseppe},
	booktitle={Proc. IEEE Int. Symp. Inf. Theory (ISIT)},
	pages={2750--2755},
	year={Taipei, Taiwan, Jun. 2023}
}
